\documentclass[twocolumn]{aastex63}
\usepackage{graphics,graphicx}
\hypersetup{urlcolor=blue}
\usepackage{natbib}
\usepackage[outdir=./]{epstopdf}
\usepackage{textcomp,gensymb}
\usepackage{xfrac}
\usepackage{xcolor}
\usepackage{multirow}

\newcommand{\vsini}{$v\sin{i_*}$}

\newcommand{\kepler}{{\it Kepler}}

\newcommand{\um}{$\mu$m}
\newcommand{\fbol}{$F_{\mathrm{bol}}$}

\newcommand{\teff}{\ensuremath{T_{\text{eff}}}}
\newcommand\kms{km~s$^{-1}$}

\newcommand{\starname}{TOI-1224}
\newcommand{\association}{MELANGE-5}
\newcommand{\age}{210}
\newcommand{\ageerr}{27}
\newcommand{\prot}{$P_{\rm{rot}}$}

\newcommand{\gaia}{\textit{Gaia}}

\usepackage[outdir=./]{epstopdf}
\usepackage{ulem}
\usepackage{mathrsfs}
\usepackage{fontawesome}
\usepackage{comment}
\usepackage[version=4]{mhchem} 

\newcommand{\tess}{\textit{TESS}}
\newcommand{\ktwo}{{\textit K2}}

\newcommand{\jwst}{\textit{JWST}}

\pdfoutput=1

\shorttitle{} 
\shortauthors{}

\begin{document}

\title{\textit{TESS} Hunt for Young and Maturing Exoplanets (THYME) X:\\
a two-planet system in the \age\, Myr \association\, Association} 

\correspondingauthor{Pa Chia Thao}
\email{pachia@live.unc.edu}

\author[0000-0001-5729-6576]{Pa Chia Thao}
\altaffiliation{NSF Graduate Research Fellow}
\altaffiliation{Jack Kent Cooke Foundation Graduate Scholar}
\affiliation{Department of Physics and Astronomy, The University of North Carolina at Chapel Hill, Chapel Hill, NC 27599, USA} 

\author[0000-0003-3654-1602]{Andrew W. Mann} 
\affiliation{Department of Physics and Astronomy, The University of North Carolina at Chapel Hill, Chapel Hill, NC 27599, USA}

\author[0000-0002-8399-472X]{Madyson G. Barber} 
\altaffiliation{NSF Graduate Research Fellow}
\affiliation{Department of Physics and Astronomy, The University of North Carolina at Chapel Hill, Chapel Hill, NC 27599, USA} 

\author[0000-0001-9811-568X]{Adam L. Kraus}
\affiliation{Department of Astronomy, The University of Texas at Austin, Austin, TX 78712, USA}

\author[0000-0003-2053-07492]{Benjamin M. Tofflemire}
\altaffiliation{51 Pegasi b Fellow}
\affiliation{Department of Astronomy, The University of Texas at Austin, Austin, TX 78712, USA}

\author[0000-0002-9446-9250]{Jonathan L. Bush}%
\affiliation{Department of Physics and Astronomy, The University of North Carolina at Chapel Hill, Chapel Hill, NC 27599, USA}

\author[0000-0001-7336-7725]{Mackenna L. Wood}%
\affiliation{Department of Physics and Astronomy, The University of North Carolina at Chapel Hill, Chapel Hill, NC 27599, USA} 
\affiliation{Department of Physics and Kavli Institute for Astrophysics and Space Research, Massachusetts Institute of Technology, Cambridge, MA 02139, USA}

\author[0000-0001-6588-9574]{Karen A.\ Collins} 
\affiliation{Center for Astrophysics \textbar \ Harvard \& Smithsonian, 60 Garden Street, Cambridge, MA 02138, USA}

\author[0000-0001-7246-5438]{Andrew Vanderburg}
\affiliation{Department of Physics and Kavli Institute for Astrophysics and Space Research, Massachusetts Institute of Technology, Cambridge, MA 02139, USA}


\author[0000-0002-8964-8377]{Samuel~N.~Quinn}
\affiliation{Center for Astrophysics \textbar \ Harvard \& Smithsonian, 60 Garden Street, Cambridge, MA 02138, USA}

\author[0000-0002-4891-3517]{George Zhou}
\affiliation{Centre for Astrophysics, University of Southern Queensland, West Street, Toowoomba, QLD 4350 Australia}

\author[0000-0003-4150-841X]{Elisabeth R. Newton} 
\affiliation{Department of Physics and Astronomy, Dartmouth College, Hanover, NH 03755, USA}

\author[0000-0002-0619-7639]{Carl Ziegler}
\affiliation{Department of Physics, Engineering and Astronomy, Stephen F. Austin State University, 1936 North St, Nacogdoches, TX 75962, USA}

\author{Nicholas Law} 
\affiliation{Department of Physics and Astronomy, The University of North Carolina at Chapel Hill, Chapel Hill, NC 27599, USA} 

\author[0000-0003-1464-9276]{Khalid Barkaoui}
\affiliation{Astrobiology Research Unit, Universit\'e de Li\`ege, 19C All\'ee du 6 Ao\^ut, 4000 Li\`ege, Belgium}
\affiliation{Department of Earth, Atmospheric and Planetary Science, Massachusetts Institute of Technology, 77 Massachusetts Avenue, Cam-
bridge, MA 02139, USA}
\affiliation{Instituto de Astrof\'isica de Canarias (IAC), Calle V\'ia L\'actea s/n, 38200, La Laguna, Tenerife, Spain}

\author[0000-0003-1572-7707]{Francisco J. Pozuelos}
\affiliation{Instituto de Astrofísica de Andalucía (IAA-CSIC), Glorieta de la Astronomía s/n, 18008 Granada, Spain}
\affiliation{Astrobiology Research Unit, Universit\'e de Li\`ege, 19C All\'ee du 6 Ao\^ut, 4000 Li\`ege, Belgium}

\author[0009-0008-2214-5039]{Mathilde Timmermans} 
\affiliation{Astrobiology Research Unit, Universit\'e de Li\`ege, 19C All\'ee du 6 Ao\^ut, 4000 Li\`ege, Belgium}

\author[0000-0003-1462-7739]{Micha\"el Gillon}
\affiliation{Astrobiology Research Unit, Universit\'e de Li\`ege, 19C All\'ee du 6 Ao\^ut, 4000 Li\`ege, Belgium}

\author[0000-0001-8923-488X]{Emmanu\"el Jehin}
\affiliation{ Space Sciences, Technologies and Astrophysics Research (STAR) Institute, Universit\'e de Li\`ege, All\'ee du 6 Ao\^ut 19C, B-4000 Li\`ege, Belgium}

\author[0000-0001-8227-1020]{Richard P. Schwarz}
\affiliation{Center for Astrophysics \textbar \ Harvard \& Smithsonian, 60 Garden Street, Cambridge, MA 02138, USA}

\author[0000-0002-4503-9705]{Tianjun~Gan}
\affil{Department of Astronomy, Tsinghua University, Beijing 100084, People's Republic of China}

\author[0000-0002-1836-3120]{Avi Shporer}
\affiliation{Department of Physics and Kavli Institute for Astrophysics and Space Research, Massachusetts Institute of Technology, Cambridge, MA 02139, USA}

\author[0000-0003-1728-0304]{Keith Horne}
\affiliation{SUPA Physics and Astronomy, University of St. Andrews, Fife, KY16 9SS Scotland, UK}

\author[0000-0003-3904-6754]{Ramotholo Sefako}  
\affiliation{South African Astronomical Observatory, P.O. Box 9, Observatory, Cape Town 7935, South Africa}

\author[0000-0002-3503-3617]{Olga Suarez}%
\affiliation{Universit\'e C\^ote d'Azur, Observatoire de la C\^ote d'Azur, CNRS, Laboratoire Lagrange, Bd de l'Observatoire, CS 34229, 06304 Nice cedex 4, France}

\author[0000-0001-5000-7292]{Djamel Mekarnia}
\affiliation{Universit\'e C\^ote d'Azur, Observatoire de la C\^ote d'Azur, CNRS, Laboratoire Lagrange, Bd de l'Observatoire, CS 34229, 06304 Nice cedex 4, France}

\author[0000-0002-7188-8428]{Tristan Guillot}
\affiliation{Universit\'e C\^ote d'Azur, Observatoire de la C\^ote d'Azur, CNRS, Laboratoire Lagrange, Bd de l'Observatoire, CS 34229, 06304 Nice cedex 4, France}

\author[0000-0002-0856-4527]{Lyu Abe}%
\affiliation{Universit\'e C\^ote d'Azur, Observatoire de la C\^ote d'Azur, CNRS, Laboratoire Lagrange, Bd de l'Observatoire, CS 34229, 06304 Nice cedex 4, France}

\author[0000-0002-5510-8751]{Amaury H.~M.~J. Triaud}%
\affiliation{School of Physics \& Astronomy, University of Birmingham, Edgbaston, Birmingham, B15 2TT, UK}

\author[0000-0002-3940-2360]{Don J. Radford}
\affiliation{Brierfield Observatory, New South Wales, Australia}

\author[0009-0006-9572-1733]{Ana Isabel Lopez Murillo}
\affiliation{Department of Physics and Astronomy, The University of North Carolina at Chapel Hill, Chapel Hill, NC 27599, USA} 

\author[0000-0003-2058-6662]{George~R.~Ricker}%
\affiliation{Department of Physics and Kavli Institute for Astrophysics and Space Research, Massachusetts Institute of Technology, Cambridge, MA 02139, USA}

\author[0000-0002-4265-047X]{Joshua~N.~Winn}
\affiliation{Department of Astrophysical Sciences, Princeton University, 4 Ivy Lane, Princeton, NJ 08544, USA}

\author[0000-0002-4715-9460]{Jon M. Jenkins}%
\affiliation{NASA Ames Research Center, Moffett Field, CA, 94035, USA}

\author[0000-0002-0514-5538]{Luke~G.~Bouma}
\affiliation{Department of Astrophysical Sciences, Princeton University, 4 Ivy Lane, Princeton, NJ 08544, USA}

\author[0000-0002-9113-7162]{Michael~Fausnaugh}
\affil{Department of Physics and Kavli Institute for Astrophysics and Space Research, Massachusetts Institute of Technology, Cambridge, MA 02139, USA}
\affil{Department of Physics \& Astronomy, Texas Tech University, Lubbock TX, 79410-1051, USA}

\author[0000-0002-5169-9427]{Natalia M. Guerrero}
\affiliation{Department of Astronomy, University of Florida, Gainesville, FL 32811, USA}

\author[0000-0001-9269-8060]{Michelle Kunimoto}
\altaffiliation{Juan Carlos Torres Postdoctoral Fellow}
\affiliation{Department of Physics and Kavli Institute for Astrophysics and Space Research, Massachusetts Institute of Technology, Cambridge, MA 02139, USA}

\begin{abstract} 
Young ($<500$\,Myr) planets are critical to studying how planets form and evolve. Among these young planetary systems, multi-planet configurations are particularly useful as they provide a means to control for variables within a system. Here, we report the discovery and characterization of a young planetary system, TOI-1224. We show that the planet-host resides within a young population we denote as \association. By employing a range of age-dating methods -- isochrone fitting, lithium abundance analysis, gyrochronology, and \gaia\, excess variability -- we estimate the age of \association\ to be \age\,$\pm$\ageerr\,\,Myr. \association\ is situated in close proximity to previously identified younger (80--110\,Myr) associations, Crius 221 and Theia 424/Volans-Carina, motivating further work to map out the group boundaries. In addition to a planet candidate detected by the \tess\, pipeline and alerted as a \tess\, Object of Interest, TOI-1224\,b, we identify a second planet, TOI-1224\,c, using custom search tools optimized for young stars (\texttt{Notch} and \texttt{LOCoR}). We find the planets are 2.10$\pm$0.09$R_\oplus$ and  2.88$\pm$0.10$R_\oplus$ and orbit their host star every 4.18 and 17.95\,days, respectively. With their bright ($K$=9.1 mag), small ($R_{*}$=0.44R$_{\odot}$), and cool (\teff=3326\,K) host star, these planets represent excellent candidates for atmospheric characterization with \jwst.

\end{abstract}


\section{Introduction}\label{sec:intro}

Young planets ($<$0.5 Gyr) offer a powerful means to study the formation and evolution of planetary systems. Similar to our own Solar System, exoplanets are expected to go through rapid evolution in the first few hundred million years after formation \citep{2012AREPS..40..251M,2001Sci...293...64A}. Throughout this early stage, planets are likely to undergo changes in their orbital \citep{Chatterjee2008}, structural \citep{owen2020constraining}, and atmospheric \citep{Oberg2011, booth2017chemical, booth2019planet} properties. Discovery and characterization of planets in this age range offers a unique window into the dynamic processes that shape planetary systems.

The \ktwo\ and \tess\ missions have enabled the discovery of transiting planets in young associations ranging from 10\,Myr \citep[e.g.,][]{Mann2016b, David2016b} to 700\,Myr \citep[e.g.,][]{Obermeier2016, Rizzuto2017, curtis2018k2}. Some similarly young planets in coeval populations have even been discovered in the \kepler\, prime field \citep{2021arXiv211214776B, Barber2022, bouma2022kepler}. The statistics of these systems have provided early evidence that young planets are larger than their older counterparts \citep{Mann2017a, Fernandes2022}. Follow-up of such systems has provided evidence that the orbits of some young, close-in planets are aligned with the equators of their host stars \citep{Zhou2020, Johnson2022} and has provided an early look at the atmospheres of young planetary systems \citep[e.g.,][]{Libby-Roberts:2020, Thao2020, Thao2023}.

Despite significant advances in recent years, the population of young transiting planets is still relatively small, falling short of the comprehensive dataset required for robust statistical analysis. The age distribution of known young planets is also heavily biased towards $<150$\,Myr and $\simeq$700\,Myr \citep{THYMEVII}. This is primarily because there are numerous nearby young groups and OB associations covering the youngest ages \citep[e.g.,][Taurus-Auriga, Sco-Cen]{2019ApJ...885L..12D, THYMEII} and several large clusters sampling the higher end \citep[Hyades and Praesepe; ][]{Mann2017a, Vanderburg2018}. The sample of 200-400\,Myr planets is drawn primarily from newly-identified associations \citep[e.g.,][]{THYMEV, Hedges2021}. 

Here, we report the discovery of a new $\simeq$200\,Myr association, \association\,, along with a two-planet transiting system orbiting one of its member stars, namely TOI-1224 or TIC 299798795. In Section~\ref{sec:obs}, we present comprehensive details of all follow-up observations of \starname\ and its parent population. In Section ~\ref{sec:analysis}, we detail the observations and analysis of \association\ association members. In Section ~\ref{sec:association}, we investigate the properties of the \association\ association, including its estimated age. In Section~\ref{sec:stellar_params}, we delve into the specific properties of the planet-hosting star, and in Section~\ref{sec:planet} we present a detailed analysis of the properties of the two planets in the system. Finally, in Section ~\ref{sec:discussion}, we provide a concise summary of our work and highlight the broader implications of these findings. We showcase their contribution to the expanding catalog of multi-planet systems and emphasize the significance of these newly discovered planets as prime candidates for future atmospheric characterization studies.

\section{Observation of \starname\ and Data Reduction}\label{sec:obs}

\begin{deluxetable}{lcccrrrrrrr} 
\tabletypesize{\footnotesize} 
\tablecaption{Time series observation log used in analysis\label{tab:obslog}}
\tablewidth{0pt}
\tablehead{
\colhead{Telescope} & 
\colhead{Filter} &
\colhead{Exp time} &
\colhead{Planet} & 
\colhead{Start Date}  \\
\colhead{} & 
\colhead{} & 
\colhead{(sec)} & 
\colhead{} & 
\colhead{(UT)}}
\startdata
    \tess\, Sector  1 & \tess\, & 120     & b, c & Jul 25 2018 \\
    \tess\, Sector 13 & \tess\, & 120     & b, c & Jun 19 2019 \\
    \tess\, Sector 27 & \tess\, & 20 & b, c & Jul 04 2020 \\
    \tess\, Sector 28 & \tess\, & 20 & b, c & Jul 30 2020 \\
    \tess\, Sector 39 & \tess\, & 20 & b, c & May 26 2021 \\
    \tess\, Sector 66 & \tess\, & 20 & b,c & Jun 02 2023 \\ 
    \tess\, Sector 67 & \tess\, & 20 & b,c & Jul 01 2023 \\ 
    \tess\, Sector 68 & \tess\, & 20 & b,c & Jul 29 2023\\ 
    \hline
    ASTEP \tablenotemark{a} & $R$ & 100 & c & Jul 4 2022 \\ 
    ASTEP & $R$ & 100 & c & Jul 22 2022 \\
    \hline
    LCO-SAAO & $g_{p}$ & 205 & b & Aug 14 2020 \\ 
    LCO-SSO  & $g_{p}$ & 240 & c & Dec 13 2022 \\ 
    \hline
    LCO-SSO  & $z_{s}$ & 55 & b & Aug 5 2020 \\ 
    LCO-SAAO & $z_{s}$ & 65 & b & Aug 14 2020 \\ 
    LCO-SAAO & $z_{s}$ & 65 & b & Sep 3 2020 \\ 
    LCO-SSO  & $z_{s}$ & 65 & c & Oct 10 2022 \\ 
    \hline
    TRAPPIST-South\tablenotemark{a} & $z$ & 25 & b & Dec 8 2020 \\ 
    TRAPPIST-South & $z$ & 35 & b & Jan 2 2021 \\  
    \hline
\enddata
\tablenotetext{a}{Only a partial transit was observed}
\end{deluxetable}

\subsection{\tess}\label{sec:tess}

TOI-1224 (TIC 299798795) was first observed by the Transiting Exoplanet Survey Satellite \citep[\tess\,;][]{Ricker2014} in Sector 1, which took place from 2018 Jul 25 to Aug 22 . Subsequently, the target was re-observed by \tess\, during Sector 13 (2019 Jun 19 to Jul 17), Sector 27 (2020 Jul 05 to Jul 30), Sector 28 (2020 Jul 31 to Aug 25), Sector 39 (2021 May 27 to Jun 24), Sector 66 (2023 Jun 02 to Jul 01), Sector 67 (2023 Jul 01 to Jul 29), and Sector 68 (2023 Jul 29 to Aug 25). For the Sector 1 and 13 data, the target was pre-selected for 120-second cadence for two guest investigator programs: G011180 (PI: C. Dressing) and G011238 (PI: S. Lepine). In the case of the later six sectors (Sector 27, 28, 39, 66, 67, 68), the target was pre-selected for 20-second cadence as part of four guest investigator programs: G03174 (PI: W. Howard), G03278 (PI: A. Mayo), G03202 (PI: R. Paudel), and G05064 (PI: W. Howard). The short-cadence observations were motivated by the star's brightness, the presence of flares, and the previous detection of a planet candidate \citep{moranta2022new}. 

We have found that the Presearch Data Conditioning Simple Aperture Photometry \citep[PDCSAP;][]{Smith2012,Stumpe2012, Stumpe2014} \tess\, light curve produced by the Science Process Operations Center \citep[SPOC;][]{Jenkins2016tess} struggles on young stars with high-amplitude stellar variability. Despite these difficulties, both planets were successfully identified by SPOC, as detailed in Section~\ref{sec:identification}. Instead, we extracted the photometry with a custom pipeline following \citet{Vanderburg2019}. This started with the Simple Aperture Photometry curves \citep[SAP;][]{Twicken2010}, which we fit with a linear model consisting of a 0.3-day basis spline, the mean and standard deviation of the spacecraft quaternion time series, seven co-trending vectors from the SPOC data conditioning, and a high-pass-filtered time series from the SPOC background aperture. Errors for each sector of data were calculated using the standard deviation of the detrended and normalized out-of-transit data. This yield errors of $\sim$0.002 and 0.005 for the 120 and 20-second cadence data. 

Since 120\,s and 20\,s cadence are both available for Sectors 27, 28, 39, 66, 67, and 68 we elected to utilize only the 20\,s data for these sectors, as the shorter cadence yields the best precision even when the data are binned back to a slower cadence \citep{Huber2022}.

Prominent flares were observed in the custom-extracted light curves \citep[also see][]{Howard2019, Gunther2020}. To mitigate their impact, we used \texttt{stella}, \footnote{\faGithub \url{https://github.com/afeinstein20/stella}}, a convolutional neural network trained for flare detection in the \tess\, short-cadence data \citep{Feinstein2020}. Utilizing the 10 models established in \cite{Feinstein2020}, we obtained an average flare prediction for each data point. Any data points with a flare probability of $>$ 80\% was excluded from the analysis. This led to the removal of 4.47\%, 3.27\%, 0.62\%, 0.26\%, 0.22\%, 0.43\%, 0.45\%, and 0.15\% of the data for Sectors 1, 13, 27, 28, 39, 66, 67, and 68, respectively. The variability in flare removal can be attributed to strong variations in the levels stellar activity and cadence differences (which impacts sensitivity to flare) between different sectors.

\subsection{ASTEP Photometry}
The Antarctica Search for Transiting ExoPlanets (ASTEP) program on the East Antarctic plateau  \citep{2015AN....336..638G,2016MNRAS.463...45M} observed three transits of planet c on the following dates: 2022 Jul 4, 2022 Jul 22, and 2022 Sep 14 (UT). The observations on Sep 14 did not yield a detectable transit signal, likely attributed to a potential transit timing variation (Section ~\ref{sec:c}); as a result, this data was excluded from the subsequent analysis. However, the remaining two observations successfully captured the transit signal. The error bars for both data set had to be adjusted to ensure that the standard deviation in the  normalized out-of-transit flux was equal to the median flux error. Scaling was applied to the data sets, with a factor of 4 for the 2022-07-4 dataset and a factor of 2 for the 2022-07-22 dataset.

The 0.4\,m telescope is equipped with two back-illuminated cameras operating in the B+V bands similar to GAIA-B (FLI Kepler KL400 sCMOS camera, $2048\times2048$ pixels), and in a red band close to the GAIA-R band (Andor iKon-L 936 CCD camera, $2048\times2048$ pixels). These cameras have an image scale of 1.05 and 1.30\,\arcsec\,pixel$^{-1}$ respectively resulting in $36\times36$ arcmin$^2$, and $44\times44$ arcmin$^2$ corrected fields of view (see \citet{schmiderSPIE2022} for further details). The fast full frame reading rate of the sCMOS sensor in the blue channel is used to guide the telescope mount at a typical rate of 0.5\,Hz, and these short exposure images are stacked to generate typical exposure times of about 1 minute. However, due to hardware signal transmission issues during the winter campaign of 2022, the blue camera could only be used during half of the season, thus leaving the red channel Andor camera as the only scientific detector for the rest of the austral winter. Exposure times and observation dates are given in Table~\ref{tab:obslog}.

Due to the low data transmission rate at the Concordia Station, the data are processed on-site using an automated IDL-based pipeline described in \citet{2013A&A...553A..49A}. The calibrated light curve is reported via email and the raw light curves of about 1,000 stars are transferred to Europe on a server in Rome, Italy, and are then available for deeper analysis. These data files contain each star’s flux computed through $10$ fixed circular aperture radii so that optimal light curves can be extracted.

\subsection{Brierfield Observatory photometry}
We observed two transits of planet b on 2019 Oct 1 and 2020 Aug 01 taken in the Johnson $I$ filter using a 0.36-meter PlaneWave CDK14 telescope at the Brierfield Observatory, located in New South Wales, Australia. The imaging system employed a Moravian 16803 camera with a pixel scale of 1.45\arcsec pixel$^{-1}$. Unfortunately, the obtained light curves did not possess the precision required for a clear detection of the transit. As a result, we decided not to include them in our transit fits.

\subsection{LCOGT photometry}

We observed a total of six transits with 1\,m telescopes in the Las Cumbres Observatory Global Telescope network \citep[LCOGT ;][]{Brown13}. We used the {\tt \tess\ Transit Finder} tool, which is a customized version of the Tapir software package \citep{Jensen2013}, to schedule the observations. These were all observed with Sinistro cameras, with a pixel scale of 0.389\arcsec\,pixel$^{-1}$. Two transits were observed using the SDSS $g'$ filter: one transit for planet b (2020-08-14) and one for planet c (2021-12-13), and and four transits were observed using the SDSS $z_s$ filter: three transits for planet b (2020-08-05, 2020-06-14, 2020-09-03) and one transit for planet c (2022-10-02). For both observations of planet c in the SDSS $g'$ and SDSS $z'$ filters, error scaling was implemented with a factor of 2. This adjustment ensured that the standard deviation in the normalized out-of-transit data equaled the median flux error value.

The images were initially calibrated by the standard LCOGT {\tt BANZAI} pipeline \citep{McCully18}. We then performed aperture photometry on all datasets using the \texttt{AstroImageJ} package \citep[AIJ;][]{Collins17}. The aperture varied based on the seeing conditions at the observatory, but we generally used a 6--10 pixel radius circular aperture for the source and an annulus with a 15--20 pixel inner radius and a 25--30 pixel outer radius for the sky background. For all observations, we centered the apertures on the source and weighted pixels within the aperture equally. All target star photometric apertures excluded flux from all known nearby \gaia\, DR3 stars. Since the event was detected on the source, the usual check of nearby sources for evidence of an eclipsing binary was not necessary. Light curves of nearby sources are available with the extracted light curves and further details on the follow-up at ExoFOP-TESS\footnote{\url{https://exofop.ipac.caltech.edu/tess/target.php?id=360156606}} \citep{EXOFOP_TESS}. Exposure times and observations dates are given in Table~\ref{tab:obslog}.

\subsection{TRAPPIST-South photometry}

We observed two transits of planet b with TRAPPIST-South \citep{Gillon2011, Jehin2011}, a 0.6m Ritchey–Chretien robotic telescope at La Silla Observatory in Chile. The TRAPPIST-South telescope is equipped with a 2K × 2K back-illuminated CCD camera with a pixel scale of 0.65 pixel$^{-1}$, resulting in a field of view of $22^{\prime} \times 22^{\prime}$. We took both transits with a Sloan-$z'$ filter.

We used the {\tt \tess\ Transit Finder} tool to schedule the observations. For data reduction and extracting photometry, we used the {\tt PROSE}\footnote{\textit{PROSE}: \url{https://github.com/lgrcia/prose}} pipeline \citep{Garcia2022}. Exposure times and observation dates are given in Table~\ref{tab:obslog}.

\subsection{SMARTS/CHIRON}\label{sec:rvs}
We observed \starname\ during four nights (2020 Nov 13, 2021 Aug 1, 2023 Feb 15, and 2023 Feb 16) with the CHIRON spectrograph on the 1.5\,m SMARTS telescope \citep{2013PASP..125.1336T}. CHIRON is a high-resolution echelle spectrograph fed by an image slicer and a fiber bundle located at Cerro Tololo Inter-American Observatory (CTIO), Chile. The observations were obtained in the low-resolution fiber mode, yielding a spectral resolution of $R\sim28,000$. We used the official CHIRON pipeline to extract the spectra as per \citet{2021AJ....162..176P}. Radial velocities are derived from a least-squares deconvolution against a non-rotating synthetic template generated from the ATLAS9 atmospheric library \citep{2004astro.ph..5087C}. We list the velocities in Table~\ref{tab:RVs}. In addition, we measure a rotational broadening of $v\sin i_\star = 22.1\pm1.2\,\mathrm{km\,s}^{-1}$ from the line broadening profile of the highest SNR spectrum.

\begin{deluxetable}{lllr}
\centering
\tabletypesize{\scriptsize}
\tablewidth{0pt}
\tablecaption{Radial Velocity Measurements of \starname\ \label{tab:RVs}}
\tablehead{\colhead{JD-2450000} & \colhead{$v$ (\kms)} & \colhead{$\sigma_v$ (\kms)} & \colhead{Instrument} }
\startdata
 9166.6767 & 14.0  & 11.3 & CHIRON \\ 
 9427.9242 & 12.41 & 0.60 & CHIRON \\ 
 9990.5292 & 11.51 & 0.49 & CHIRON \\ 
 9991.5496 & 12.03 & 0.66 & CHIRON \\ 
\enddata
\end{deluxetable}

\subsection{Speckle Imaging}\label{sec:speckle}

\begin{figure}[ht]
    \centering
    \includegraphics[width=0.45\textwidth]{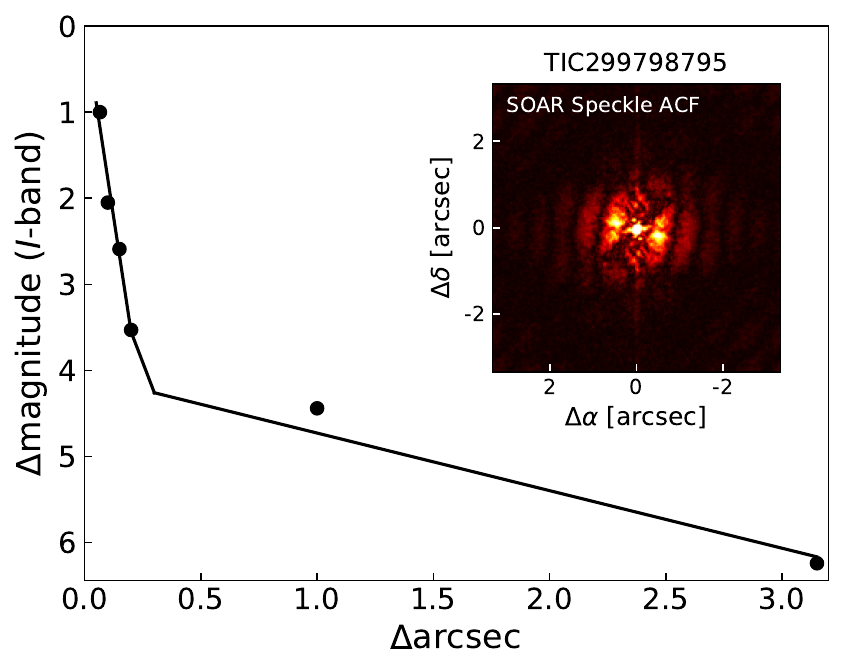} 
    \caption{Detection limits from the speckle images of \starname. The sub-panel in the top right shows the narrow-band 8320\AA, reconstructed image.
    \label{fig:speckle}}
\end{figure}

We observed \starname\ on 10 February 2020 from the 4.1-m Southern Astrophysical Research (SOAR) telescope with speckle interferometry in the \textit{I}-band \citep{Tok2018b}. We took these observations following general observing strategy for \tess\ targets as described in \citet{Ziegler2020}. This yielded an estimated contrast limit of $\Delta I$ = 6.2 at 1\arcsec\ (Figure~\ref{fig:speckle}). We detected no companions.

\section{Observations and analysis of \association\ candidate members} \label{sec:analysis}
Our motivation for observing the candidate members of \association\ was threefold: to confirm the existence of the association, ascertain that it constitutes a single-aged population, and precisely measure its age. We also sought to explore the relationship between this group and nearby ones identified in \citet{Moranta2022}. 

\subsection{Target Selection and identification of \association}\label{sec:target}

We initially searched for a candidate association around \starname\, using the open source code, \texttt{Comove} \footnote{\faGithub \href{https://github.com/adamkraus/Comove}{ https://github.com/adamkraus/Comove}}, which is described in detail in \citet{THYMEV}. To briefly summarize, \texttt{Comove} utilizes the astrometric data from \gaia\, Data Release 3 \citep[DR3;][]{Lindegren_2021} to identify stars that exhibit potential co-moving characteristics within a specified spatial and kinematic range. It estimates the 3D distance and expected tangential velocity ($V_{off}$) of all stars within the defined radius assuming a $UVW$ matching \starname\, and utilizing radial velocity data from \gaia\ DR3 and archival sources.

We initially selected stars with a tangential velocity difference of $<5$\kms\ and a position $<$50\,pc of \starname. The resulting target list was heavily contaminated by unassociated stars, as evident from a large spread in the color-magnitude diagram (CMD) and radial velocities, as well as a significant population of cool white dwarfs. As we discuss further in Section~\ref{sec:crius}, a significant portion of the contamination comes from a nearby younger population (which could bias our age estimates). Thus, we opted for a tighter cut of $3$\kms\ and 35\,pc. This yielded a list of 159 stars. 

\begin{figure*}[ht]
    \centering
    \includegraphics[width=0.47\textwidth]{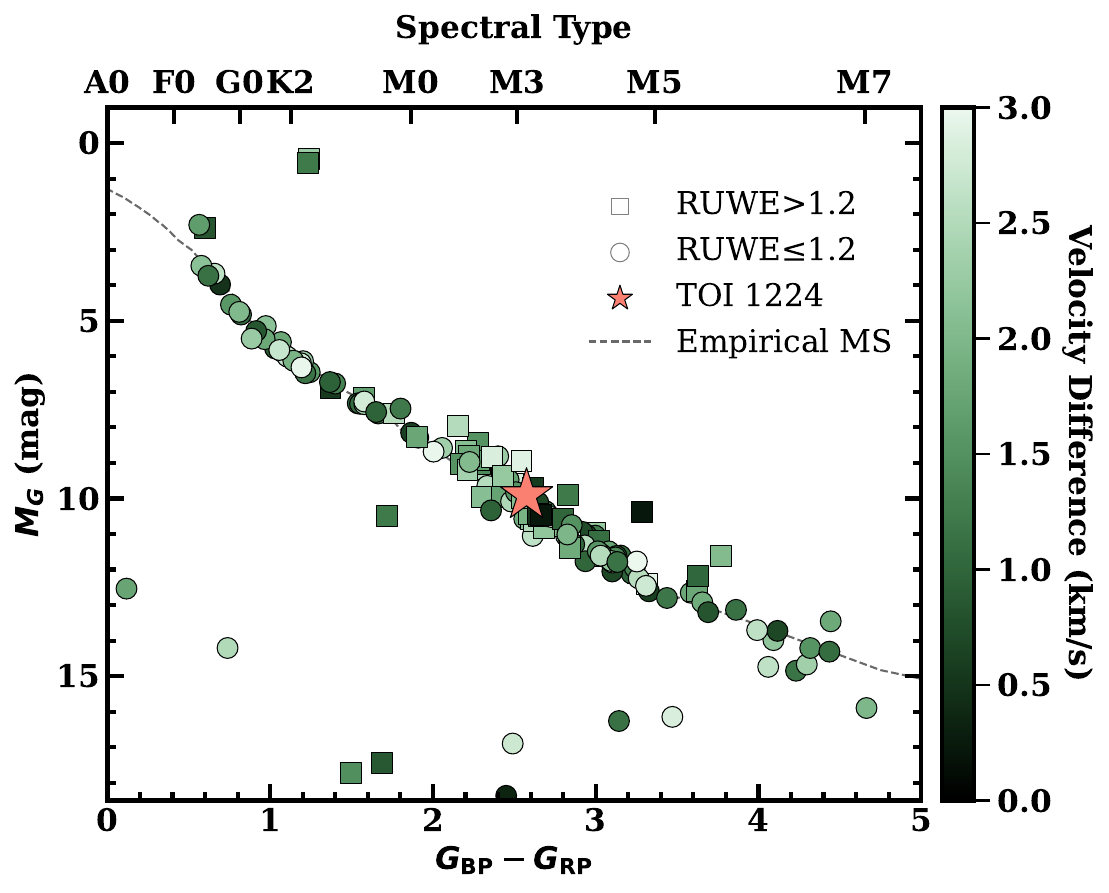} 
    \includegraphics[width=0.47\textwidth]{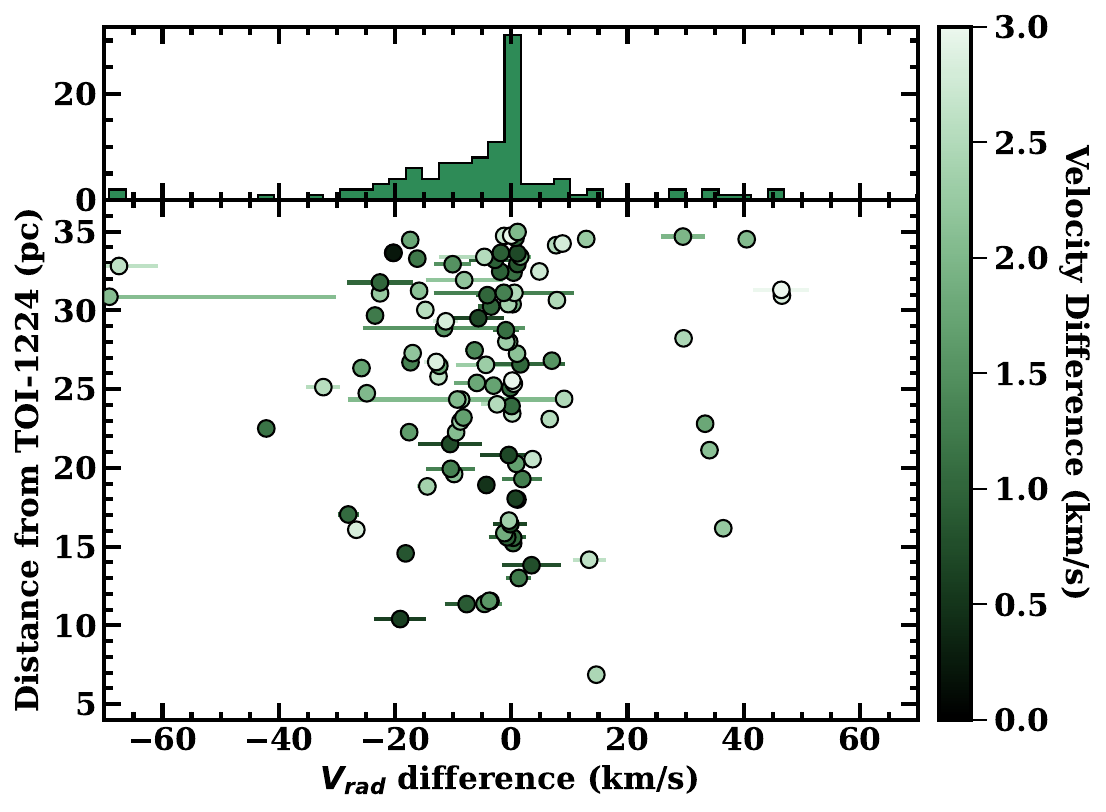} 
    \caption{\textit{Left:} Color-magnitude diagram of sources within 35\,pc and 3\kms\ of \starname. Points are color-coded by their velocity difference, and TOI-1224 is denoted as a pink star. Targets with a high RUWE (more likely to be binaries) are denoted as squares. \textit{Right: } Radial velocities of candidate members of \association\ compared to the value predicted assuming they have identical $UVW$ to \starname. RV data is extracted from \gaia\, DR3 and archival sources. The excess of points around zero is far above what is possible by chance and is far narrower than the velocity distribution of all Solar neighborhood stars. Stars closer to zero radial velocity difference are also statistically closer in tangential velocity. The excess of sources below (but near) zero may be due to uncertainties in \starname's velocity, \starname\ landing slightly off-center of the association, or coincident alignment with other groups (see Section~\ref{sec:association}) or Galactic motions
    \label{fig:gaia_cmd}}
\end{figure*}

The resulting list formed a relatively tight CMD (Figure~\ref{fig:gaia_cmd}) consistent with a single-aged main-sequence population and a handful of outliers (a mix of non-members and poorer photometry/astrometry). Selected stars also cluster around the expected central radial velocity (again with some clear outliers: Figure~\ref{fig:gaia_cmd}). Since radial velocity and CMD information were not used to select members, these are compelling pieces of evidence that the grouping represents a real association. We refer to this association as \association\ (Membership and Evolution by Leveraging Adjacent Neighbors in a Genuine Ensemble), following the scheme from earlier papers \citep{THYMEV}.

The selection above is over-inclusive, in that there are obvious outliers in kinematic, spatial, and CMD space. Additional down-selections, such as enforcing a radial velocity cut, would likely yield a cleaner list. However, we have adopted a more inclusive list to start, and perform additional cuts as needed to better determine the age in Section~\ref{sec:age}.  

\subsection{Spectroscopy}
\subsubsection{LCO/NRES}

We obtained high-resolution ($R\simeq53,000$) optical (3800-8600\AA) spectra of candidate association members using the LCO Network of Robotic Echelle Spectrographs \citep[NRES;][]{NRES2018}. We prioritized targets with $G<9$ to ensure SNR$>20$ around the 6707\AA\, Lithium line. In total, we obtained spectra for 5 targets this way.

All NRES data were reduced and wavelength calibrated using the BANZAI-NRES pipeline\footnote{\faGithub  \url{https://github.com/LCOGT/banzai-nres}} \citep{McCully2022}.

\subsubsection{Archival HARPS Spectra}
We downloaded spectra from the ESO science archive taken with the High Accuracy Radial velocity Planet Searcher \citep[HARPS;][]{Mayor2003} fiber-fed Echelle Spectrograph at the ESO La Silla 3.6m telescope. HARPS spectra cover a spectral range of 3780–6910\AA\ at high resolution ($R\simeq$115,000). Archival spectra were taken as early as 2014 and as late as 2020. We required a SNR of at least 30 around the 6707\AA\ Li line. In the case of multiple spectra for the same target, we use the one with the highest SNR. In total, we retrieved spectra for 15 candidate members of \association. 

The ESO archive provided HARPS data reduced with the standard online HARPS data reduction pipeline. 

\subsubsection{Li Equivalent Widths}\label{sec:li_eqw}

Lithium (Li) is destroyed in the interiors of stars. Mixing can therefore deplete lithium levels in the photosphere, creating a relation between age and the measured lithium abundance. For fully convective stars, total Li depletion is rapid. For Sun-like stars, Li depletion depends on the mixing timescale and the depth of the convective regions. Additional effects, like rotation, complicate this relation and add significant scatter even for single-aged populations \citep{Somers2015}. Still, the lithium levels in a range of stars within a cluster can provide a useful constraint on the association's age. 

We measured the equivalent width of the 6708\AA\ Li line from the HARPS and NRES spectra (Section~\ref{sec:obs}). We first placed the reduced spectra into the star's rest frame by cross-correlating against a matching PHOENIX model \citep{Husser2013}. We then performed a least-squares fit to the region around the Li line (6706\AA--6710\AA) assuming a Gaussian. We removed other spectral lines in the region using a 5$\sigma$ iterative clipping. We used the estimate of the continuum level and line width from the Gaussian to calculate the equivalent width. 

For nearly all stars, the spectral resolution was sufficient to separate out the nearby iron line. For those with visible contamination, we fit a double-Gaussian and subtracted the iron line model before computing the equivalent width. 

We estimated the uncertainties on our equivalent widths using a Monte Carlo (MC) approach. Specifically, we perturbed the spectra, refit the line, then recomputed the equivalent width 10,000 times and used the scatter in the resulting values as a proxy for the uncertainty. Comparison to equivalent widths from \citet{THYMEV}, \citet{Barber2022}, and \citet{THYMEVII} suggest these uncertainties are missing a systematic error term of $\simeq10$mA, which we added in on top of the MC uncertainties. The same analysis (MC and comparison to existing measurements) indicated we are insensitive to equivalent widths below 20$m\AA$; we set all non-detections to this limit. 

We drew two additional lithium measurements from \citet{Torres2008}. These equivalent widths were estimated using similar high-resolution spectra and comparison between their targets and those with NRES or HARPS data showed a negligible offset and comparable uncertainties. The results of our detailed abundance measurements are detailed in Table ~\ref{tab:toi1224_friends}.

\subsection{\tess\ photometry}\label{sec:unpopular}

We extract \tess\ light curves for candidate association members using Causal Pixel Models \citep[CPM;][]{Wang2016} run on the full-frame images (FFIs). Specifically, we used the \texttt{unpopular} package\footnote{\faGithub \url{https://github.com/soichiro-hattori/unpopular}} to build a model of systematics \citep{Hattori2021}. The details of the \texttt{unpopular} extraction used here are reported in \citet{Barber2022}, \citet{Wood2023}, and \citet{Vowell2023}, each of which were applied to measuring rotation periods in young associations.

\subsection{Rotation Periods}

To ensure that we get high-quality rotation periods, we first removed those stars with $T>$15 mag and/or a flux contamination ratio $>$1.5. 

To estimate the rotation period of each target we employed the Lomb-Scargle \citep[LS;][]{LombScargle} algorithm using the fast implementation \citep{Press1989} in \texttt{astropy}. For each sector of \tess\ photometry, we searched over a period range of 0.1-15\,days, adopting the peak in the LS periodogram as a candidate period. Across sectors, we selected the period corresponding to the highest LS power. To confirm these measurements, we phase-folded the single-quarter light curves to the candidate period and examined the signals' consistency across quarters. We performed an eye-check in the style of \citet{Rampalli2021}, labeling obvious rotations as Q0, questionable rotations as Q1, spurious detections as Q2, and non-detections as Q3.

We retained only Q0 and Q1 periods. In total, 87 stars passed this cut. The remainder is a mix of non-members and targets too faint for reliable extraction of the \tess\ light curve. A summary of each period and quality assessment can be found in Table ~\ref{tab:toi1224_friends}.

\section{Properties of the \association\, Association }\label{sec:association}

\subsection{Age}

\subsubsection{Lithium-based age}\label{sec:lithium}

\begin{figure}[ht]
    \centering
    \includegraphics[width=0.45\textwidth]{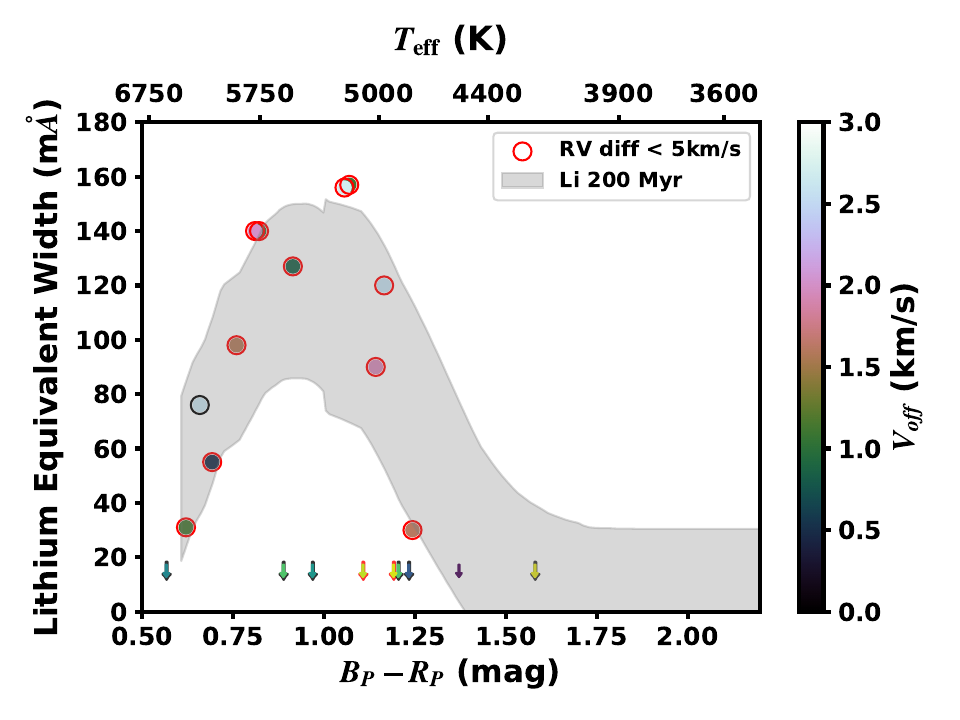} 
    \caption{Lithium equivalent widths of candidate members of \association\ as a function of \gaia\ color. The top axis shows approximate \teff\ for reference. Points are color-coded based on their radial velocity. Non- or weak detections are shown as arrows. Targets with radial velocities within 5 km\,s$^{-1}$ of the predicted value are outlined in red. The predicted 200\,Myr sequence from \texttt{EAGLES} is shaded. 
    \label{fig:lithium}}
\end{figure}

We show the lithium equivalent widths (EW) of potential members in Figure~\ref{fig:lithium}. We converted these measurements into an age estimate of the association using the \texttt{EAGLES}\footnote{\faGithub  \url{https://github.com/robdjeff/eagles}} software \citep{Jeffries2023}. \texttt{EAGLES} was calibrated using spectroscopy from the \gaia-ESO survey of clusters with well-established ages \citep[e.g.,][]{Randich2022,Jackson2022}.

The \texttt{EAGLES} code is not designed to work with non-members included, so we only included targets with radial velocities within 5\,km\,s$^{-1}$ of the predicted value. This left 14 stars. \texttt{EAGLES} works in \teff, so we converted the observed \textit{Gaia} $B_P-R_P$ colors to \teff\, using the \citet{Pecaut2013} empirical tables\footnote{\url{https://www.pas.rochester.edu/~emamajek/EEM_dwarf_UBVIJHK_colors_Teff.txt}}. 

We ran \texttt{EAGLES} in `cluster' mode, including uncertainties on \teff\ and lithium, which yielded an age of $177^{+45}_{-36}$\,Myr. There are two stars in the sample with no detected lithium even though a detection would be expected at this age. Both stars land near the tangential velocity cutoff for selecting members (Section~\ref{sec:target}), suggesting they might not be members. Removing these two stars resulted in an age of $185\pm35$\,Myr -- an insignificant effect on the age and a small reduction on the upper uncertainty. Either age was consistent with all other determinations, although our isochrone fit rules out the lower end of both posteriors ($<150$\,Myr).

\subsubsection{Gyrochrone age}\label{sec:rotation}
\begin{figure}[ht]
  \centering
  \includegraphics[width=0.48\textwidth]{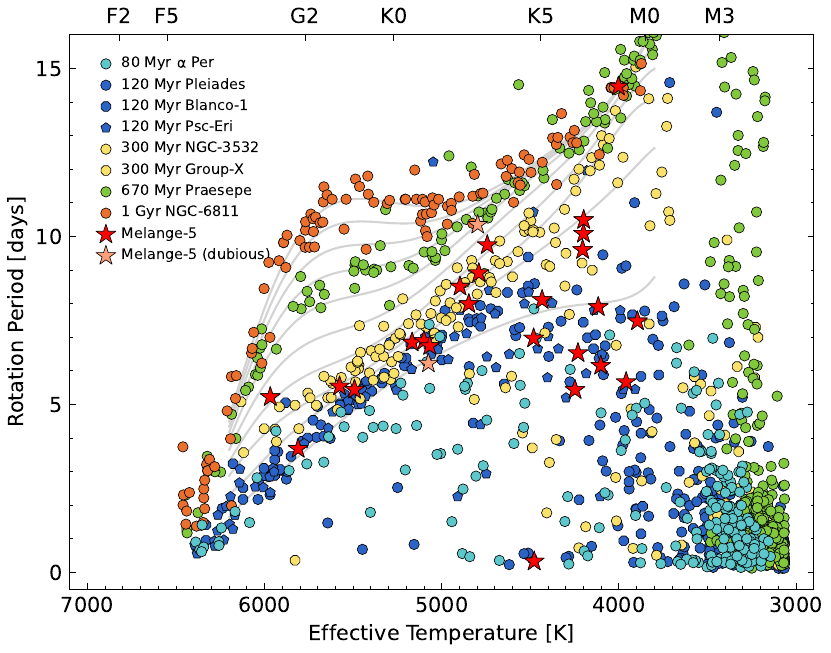}
  \caption{Rotation periods of MELANGE-5 compared to rotation periods of
  benchmark open clusters.  MELANGE-5 members are represented by red stars and the shade indicates the quality of the rotation, with darker red (0) representing a clearly observed rotation in the light curve and lighter red (1) representing a questionable period. MELANGE-5 falls above the 120\,Myr clusters, and slightly below the 300\,Myr clusters.  The gray lines are the ``mean'' empirical isochrones fitted by \citet{Bouma_2023} with ascending time intervals of 100\,Myr, 200\,Myr, and so on.  The astrophysical scatter about these mean lines sets the empirical precision limit for this age-dating method.}
  \label{fig:prot}
\end{figure}  

We show the \prot-color sequence for \association\ alongside benchmark open clusters younger in Figure~\ref{fig:prot}.  The rotation periods for these clusters are from $\alpha$~Per \citep{Boyle_2023}; the Pleiades
\citep{Rebull_2016}; Blanco-1 \citep{Gillen_2020}; Psc-Eri \citep{curtis_2019}; NGC-3532 \citep{Fritzewski_2021}; Group-X \citep{Messina_2022}; Praesepe \citep{Douglas_2019}; and NGC-6811 \citep{Curtis_2019_ngc6811}.  To generate the plot, we used the effective temperature scale from \citet{Curtis_2020}, the same extinction corrections and intrinsic age scale for these clusters as
listed in Table~1 of \citet{Bouma_2023}, and the data behind the relevant
figure from \citet{Bouma_2023}.  The most relevant assumed ages are those for the Pleiades \citep{GalindoGuil_2022} and NGC-3532 \citep{Fritzewski_2019}.  We assumed negligible reddening for \association\ due to its proximity, and only plot the stars in the association with $3800 < T_{\rm eff} / {\rm K} < 6200$, which are most diagnostic for the rotation-based age.

Visually, Figure~\ref{fig:prot} shows that on average, the association of \association\ falls above the 120\,Myr clusters, and slightly below the 300\,Myr clusters.  For any individual star, this measurement would not be particularly significant.  However, the ensemble of stars falling above the 120\,Myr sequence is highly significant because almost every $\approx$G2-K4 star falls above it, and this is the regime with the most diagnostic power.  To leverage the ensemble information, we use the hierarchical Bayesian framework implemented in ${\tt gyro-interp}$, and discussed by \citet{Bouma_2023}.  This model assumes that the intrinsic age of any star in the association is drawn from a Gaussian with an unknown mean and intrinsic scatter.  We considered two subsets of the rotation data when performing the inference: {\it i)} all stars with $3800 < T_{\rm eff} / {\rm K} < 6200$, and {\it ii)} only the ``good'' rotation periods.  These assumptions yielded a mean cluster age and associated $1-\sigma$ uncertainties of $t$=$283^{+92}_{-59}$\,Myr when considering all rotation periods, and $234^{+77}_{-56}$\,Myr when only considering the ``good'' periods.  Both estimates agree with the visual impression from Figure~\ref{fig:prot} that \association\ is between the 120\,Myr and 300\,Myr calibration sequences.  The statistical uncertainty on this statement is limited by the intrinsic astrophysical scatter in the cluster rotation sequences at these ages.

\subsubsection{Isochronal Age}\label{sec:isochronal_age}

We compared \gaia\ photometry of candidate \association\ members to predictions from the PARSECv1.2 isochrones \citep{PARSEC}. We used the mixture model described in the Appendix of \citet{THYMEVI}, which follows the statistical methods from \citet{HoggRecipes}. To summarize, the model is a mixture of two models. The first represents a single-aged single-star population, and is drawn from the isochrones as a function of by age ($\tau$) and reddening ($E(B-V)$). The second is meant to capture outliers (mostly non-members), and is described by an offset from the isochrone model ($Y_B$) and a Gaussian variance around that offset ($V_B$). The final parameter ($f$) captured missing uncertainties such as variation in reddening between members, model systematics, and underestimated uncertainties in the \gaia\, data.

For our comparison, we excluded stars with RUWE$>$1.5 \citep{Ziegler2020,Wood2021}, those with \gaia\ photometry or parallax SNR $<$20, and those outside the model grid photometry (including white dwarf candidates). The mixture model can handle some of the stars eliminated by these cuts by calling them outliers, but a less complex outlier population tends to provide better fits (e.g., including white dwarfs biases the outlier distribution). The fit was run using \texttt{emcee} \citep{Foreman-Mackey2013} with 30 walkers for 20,000 steps, which was more than sufficient for convergence. 

\begin{figure}[ht]
    \centering
    \includegraphics[width=0.47\textwidth]{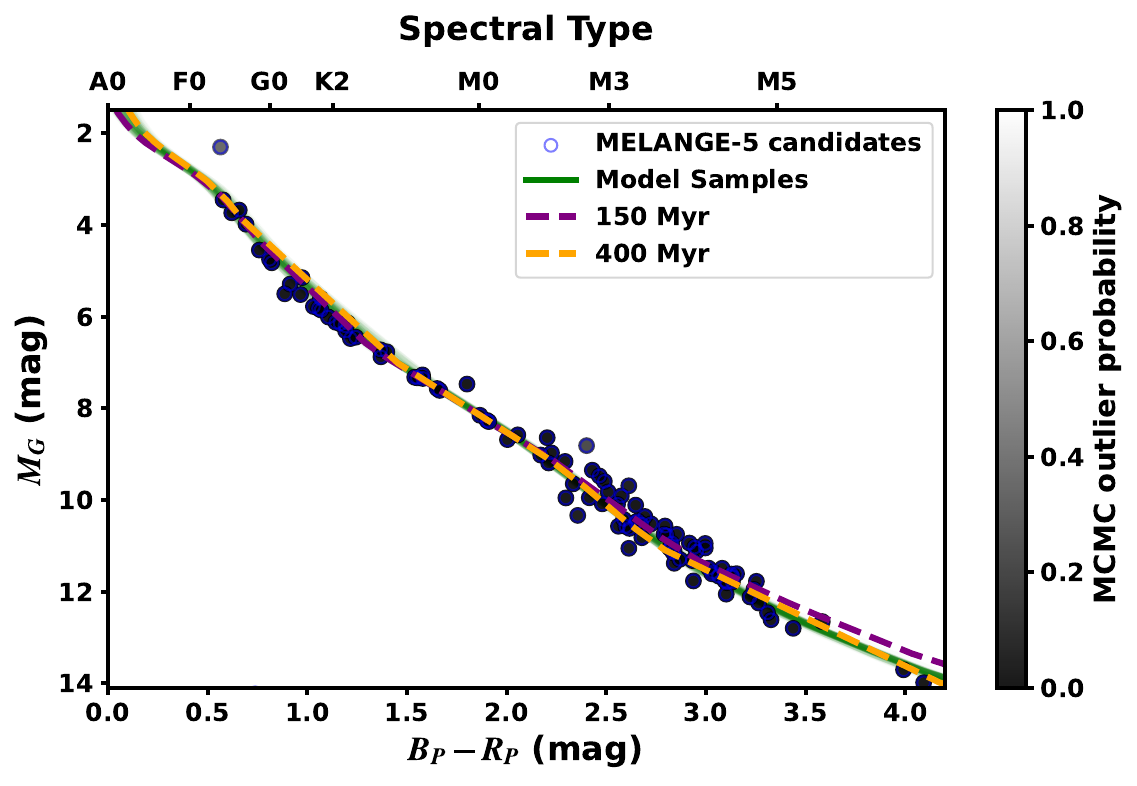} \caption{Example CMD of \association\ (black circles) fit using a mixture model. Each point is shaded by the probability of being part of the primary population (the single-star single-aged group). The green lines are 100 random draws from the MCMC posterior. For reference, we included 400\,Myr and 150\,Myr isochrones.} \label{fig:isochrones}
\end{figure}

As shown in Figure~\ref{fig:isochrones}, the CMD fit provides only large bounds on the age of \association. The minimum age is set by the lack of pre-main-sequence M dwarfs ($\simeq$150\,Myr for cooler than M5) and the maximum age by the lack of post-main-sequence evolution in the F stars (approximately 800\,Myr). The younger limit is set primarily by just 4 stars that sit below the 150\,Myr sequence.

\subsubsection{Age from Gaia excess variability}\label{sec:gaiavar}

\cite{Barber2023} presented a method to estimate the age of an association from the excess variance in \gaia\ photometry \citep{Riello2021_phot}. The method uses a Skumanich-like relation between stellar activity and age and takes advantage of the fact that more variable sources will have higher photometric uncertainties than their quiet counterparts. The method considers the distribution of variability over a population of co-eval stars, thereby averaging over stellar inclinations and star-to-star variability. Using the open source code, \texttt{EVA}\footnote{\faGithub \href{https://github.com/madysonb/EVA}{  https://github.com/madysonb/EVA}} (Excess Variability-based Age), this technique yielded an age of $176_{-51}^{+98}$\,Myr, which is consistent with our other age determination methods. 

\begin{deluxetable}{lr}
\tablecaption{Age determinations
\label{tab:ages}}
\tablecolumns{1}
\tablewidth{0pt}
\tablehead{
\colhead{Method} & 
\colhead{Age (Myr)}} 
\startdata
Lithium (\texttt{Eagles}) & $185\pm35$\\
Isochrone (PARSEC) & $150-800$ \\
Gyrochronology &  $283^{+92}_{-59}$\\
Variability  & $176_{-51}^{+98}$ \\
\hline
\textbf{Adopted} & $210\pm$27 \\ 
\enddata 
\end{deluxetable}

\subsubsection{Combined Age}\label{sec:age}

Our four methods for estimating the age of \association\ are mostly independent of each other, in that they rely on different datasets and approaches. They do rely on a common set of members, but each method is relatively robust to the exact membership list. Gyrochronology, lithium, and \gaia\ excess variability are also calibrated on a similar set of clusters, although the underlying cluster population ages are known more precisely than the determinations here. Hence, we can combine these ages into one more  precise age.

We find an adopted age by maximum likelihood, assuming (assymetric when required) Gaussians except for the isochrone age, for which we assume a uniform distribution from 150-800\,Myr. We estimate the age for MELANGE-5 to be 210$\pm$27\,Myr. We adopt this value for the age of the group. A summary of each age determination can be found in Table~\ref{tab:ages}.

\subsection{Membership List Contamination Rate}

Following \citet{Barber2022}, we can estimate the fraction of interlopers using the \gaia\ radial velocities. As explained in Section~\ref{sec:target}, the target list was selected based on position and tangential velocity. Since the probability of radial velocity alignment is small, the fraction of stars with inconsistent velocities is a rough estimate of the fraction of stars that are field contaminants. However, tight binaries (which may have discrepant velocities) could be falsely assigned as non-members, increasing the contamination rate, and field stars that align with the group by chance could be falsely flagged as members, decreasing the contamination rate.

We assumed all stars with 3$\sigma$ consistent radial velocities are members and adopt an internal velocity dispersion of 0-3\kms\ along the line of sight. This yielded a contamination fraction of 38-55\% (the high value corresponding to the lowest assumed velocity dispersion). Assuming that \starname\ is 2\kms\ off from the true group center (as suggested by Figure~\ref{fig:gaia_cmd}) decreases the contamination to 28-40\%. 

The lithium suggests a similar contamination fraction. Focusing on those stars where we expect a lithium detection (i.e., $0.6<B_P-R_P<1.3$), there were 18 stars with data, and 12 of them (67\%) showed elevated lithium levels consistent with the age of the group (see Figure~\ref{fig:lithium}), for a contamination rate of 33\%. As with radial velocities, this may include non-members with elevated lithium by chance, and exclude some members that are depleted due to astrophysical variation \citep{Somers2017}. 

The rotation periods yield a similar estimate. Of the 159 candidate members, 34 of them were too faint or had contamination too high for a useful \tess\ light curve. Of the remaining, 81 stars had rotation periods consistent with membership. Most of the rest had no clear rotation (Q3 or Q4; 38), with 6 Q0 or Q1 rotators landing clearly above the color-period sequence (Figure~\ref{fig:prot}). This gives a contamination fraction of 35\%. Using rotation periods in this way has similar issues to lithium and velocities; some non-members will match by chance, some members will have no apparent rotation because they are pole-on, and some rotation periods may appear discrepant or match because we actually measured an alias of the true period. 

Overall, these three methods yield a similar result, suggesting that about 35\% of the stars in our target list are field stars. All three calculations may be underestimates due to contamination from the nearby 90\,Myr population(s) (Section~\ref{sec:crius}). These stars will be closer in radial velocity than random field stars, rotate faster, and have elevated lithium levels. However, the lack of pre-main-sequence M dwarfs in the \association\ CMD suggests contamination from these younger groups is relatively low.

\subsection{Relation to Crius 221, Volans-Carina, and Theia 424}\label{sec:crius}

\begin{figure*}[ht]
    \centering
    \includegraphics[width=1.\textwidth]{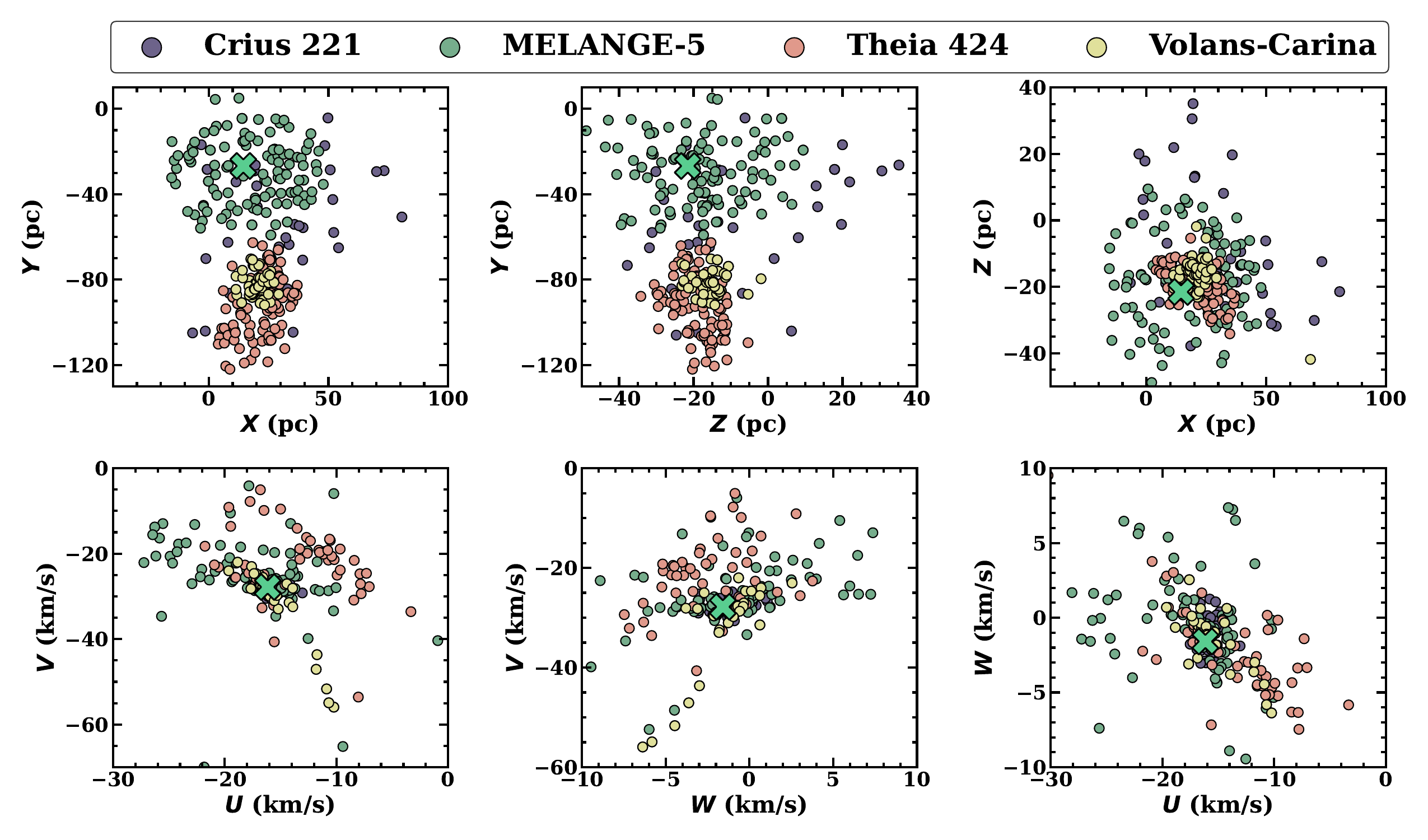} \caption{The spatial spread and velocity spread of of MELANGE-5 (green) compared to the nearby associations, Crius 221 (purple), Theia 424 (pink), and Volans Carina (yellow). The top row shows the galactic position ($X$, $Y$, and $Z$) and the bottom row shows the velocities ($U$, $V$, and $W$) for each population. MELANGE-5 ($\sim$ 200 Myr) overlaps with Crius 221 ($\sim$100 Myr), but does not intersect with Theia 424 or Volans Carina. The position of TOI 1224 is marked with green X.} \label{fig:other_associations}
\end{figure*}

\citet{Moranta2022} include \starname\ in the list of Crius 221 members\footnote{They specifically noted the planet host (TOI) as a planet in their newly-identified group} and also suggest that Crius 221 is part of Volans-Carina \citep{Gagne2018}, Theia 424 \citep{KounkelCovey2020}, and Oh 30 \citep{Oh2017}. Indeed, all of these groups (and \association) show significant spatial and kinematic overlap (Figure~\ref{fig:other_associations}). This raises the question if MELANGE-5 is a unique group or an extension of these known populations. 

\begin{figure}[ht]
    \centering
    \includegraphics[width=0.47\textwidth]{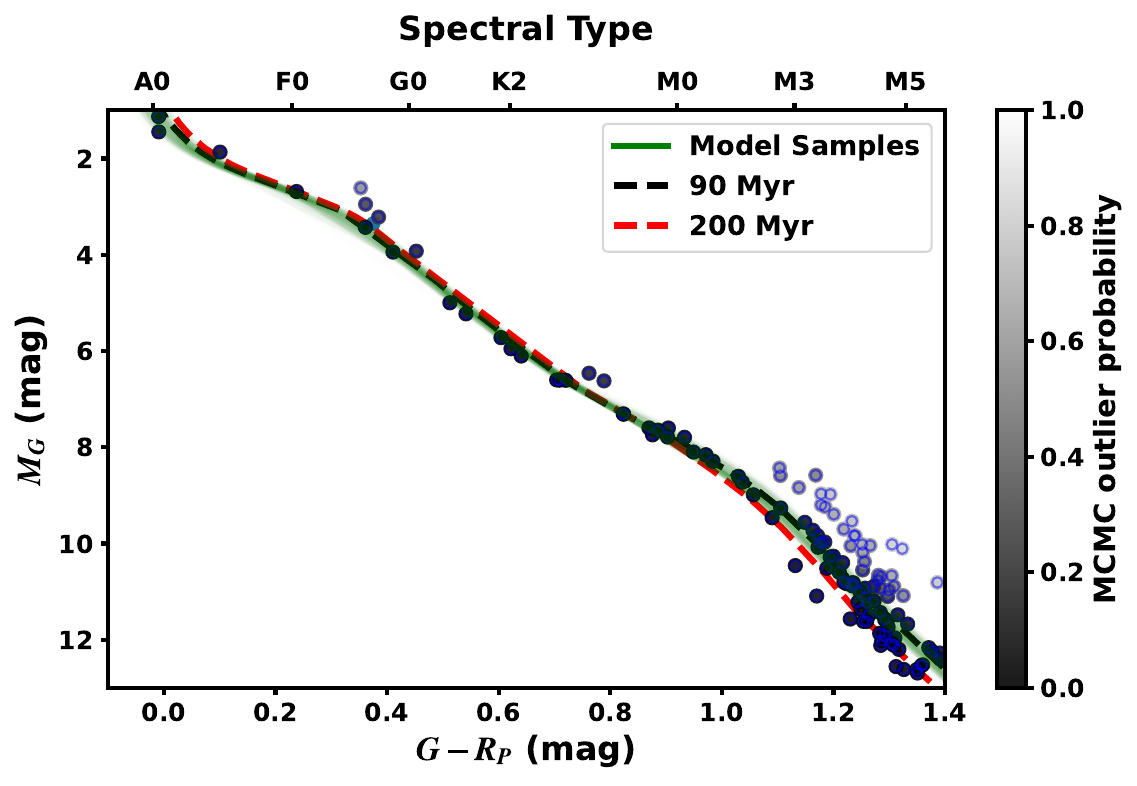} \caption{CMD of members of Theia 424 from \citet{KounkelCovey2020} compared to PARSECv1.2 solar metallicity models using a mixture model (Section~\ref{sec:isochronal_age}). The green lines are 100 random draws from the fit posterior and individual points are shaded by the probability that they are part of the main (single-star single-aged) population. The group is well described by a $\simeq$90\,Myr isochrone (black dashed line) with a visible binary sequence above and a small number of field interlopers below.} \label{fig:theia424}
\end{figure}

If all groups are part of a common parent population they should have the same age. However, \citet{Gagne2018} found that Volans-Carina is only $89^{+5}_{-7}$\,Myr, which is inconsistent with our age estimate for MELANGE-5. As an additional check, we re-ran our isochronal fit described in Section~\ref{sec:isochronal_age} on Crius 221, Volans-Carina, and Theia 424 (Oh 30 had too few members to test). For consistency, the method was unchanged from our analysis of MELANGE-5. Using the PARSEC models, we measured ages of 98$^{+23}_{-9}$\,Myr, 93$\pm$10\,Myr, and 99$^{+7}_{-12}$\,Myr for Crius 221, Volans-Carina, and Theia 424, respectively. Crius 221 only has two stars passing our selection criteria later than M3 (where we expect pre-main-sequence stars at this age); if both are binaries, the age could be significantly older. However, the assigned age of the other two are more robust to the target list. Theia 424 in particular, matches the 90-100\,Myr isochrone, has a clearly separated binary sequence (Figure~\ref{fig:theia424}), and includes a smaller population of likely main-sequence mid-M interlopers. 

We conclude that although these groups overlap with MELANGE-5, they are likely distinct populations. 

\subsection{Membership of TOI-1224}

A separate question is the membership of \starname. The target is unambiguously young, but there is some ambiguity about which group it resides within. \starname\ is spatially separate from Theia-424/Volans-Carina (Figure~\ref{fig:speckle}) and is not listed as a member of either. It is, however, listed as a member of Crius 221. By design, \starname\ is at the center of \association. Selecting any of the other high-probability members (stars showing high lithium and rotation) and re-running \texttt{Comove} includes \starname\ as a nearby member. 

Most likely, \starname\ is part of either Crius 221 or \association. \citet{Moranta2022} refers Crius 221 as a corona of Volans-Carina; however, it is quite diffuse compared to the other groups considered here. It may be some mix of Volans-Carina and \association. Assuming Crius 221 is a distinct association, the strongest piece of evidence that \starname\ is a member is the target's CMD position, which is marginally better match to a 90\,Myr isochrone than 200\,Myr. The difference is small and is consistent with intrinsic scatter in the CMD (Section~\ref{sec:isochronal_age}). The mixture model finds a 99.4\% chance that \starname\ is part of the main population. Although this is misleading because it only compares the probability that the star is part of \association\ to that of non-members (mostly the field). Indeed, a fit to the Crius-221 group using the same code finds a similar (99.1\%) probability. As an M dwarf, the lack of lithium and fast rotation period for \starname\ are consistent with both 90\,Myr and 200\,Myr. Thus, we cannot rule out that \starname\ is part of Crius 221, but the kinematics and position favor membership with \association. We briefly discuss in Section~\ref{sec:discussion} what impact this has on the system parameters.

\section{Properties of TOI 1224}\label{sec:stellar_params}

\subsection{$M_*$ and $R_*$ from empirical relations}\label{sec:emp}

 We estimated stellar mass and radius using the empirical $M_K-R_*$ and $M_K-M_*$ relations from \citet{Mann2015b} and \citet{Mann2019a}\footnote{\faGithub \href{https://github.com/awmann/M_-M_K-}{  https://github.com/awmann/M\_-M\_K-}}, respectively. We used the relations without metallicity as a parameter, although using [Fe/H]=0 yielded consistent results. For $M_{K_S}$, we used the parallax from \gaia\ DR3 and $K_S$ photometry from the Two Micron All Sky Survey \citep[2MASS; ][]{Skrutskie2006}. We ignored the effects of extinction, as the star is within the local bubble. This yielded a radius of $0.404\pm0.012R_\oplus$ and a mass of $0.400\pm0.010M_\oplus$.
 
 Uncertainties in $M_*$ and $R_*$ account for both measurement errors (in $K_S$ and parallax) and uncertainties in the calibrations. While high activity may change the inferred or true parameters of stars \citep[e.g.,][]{Feiden2012b}, there is no evidence that the $M_K-R_*$ and $M_K-M_*$ relations used here are inaccurate for active main-sequence stars  \citep{Mann2015b, Mann2019a}. However, the age of \starname\ puts it near the zero-age main-sequence, suggesting it could be pre-main-sequence depending on the model grid and assumed age. Correcting for this would yield a lower mass and larger radii, but the change is expected to be comparable to the measurement uncertainties in both cases. As an additional check, we estimated $R_*$ from the spectral energy distribution (SED) below. 

\subsection{\teff, $L_{*}$ and $R_{*}$ from the SED}

\begin{figure}[ht]
    \centering
    \includegraphics[width=0.45\textwidth]{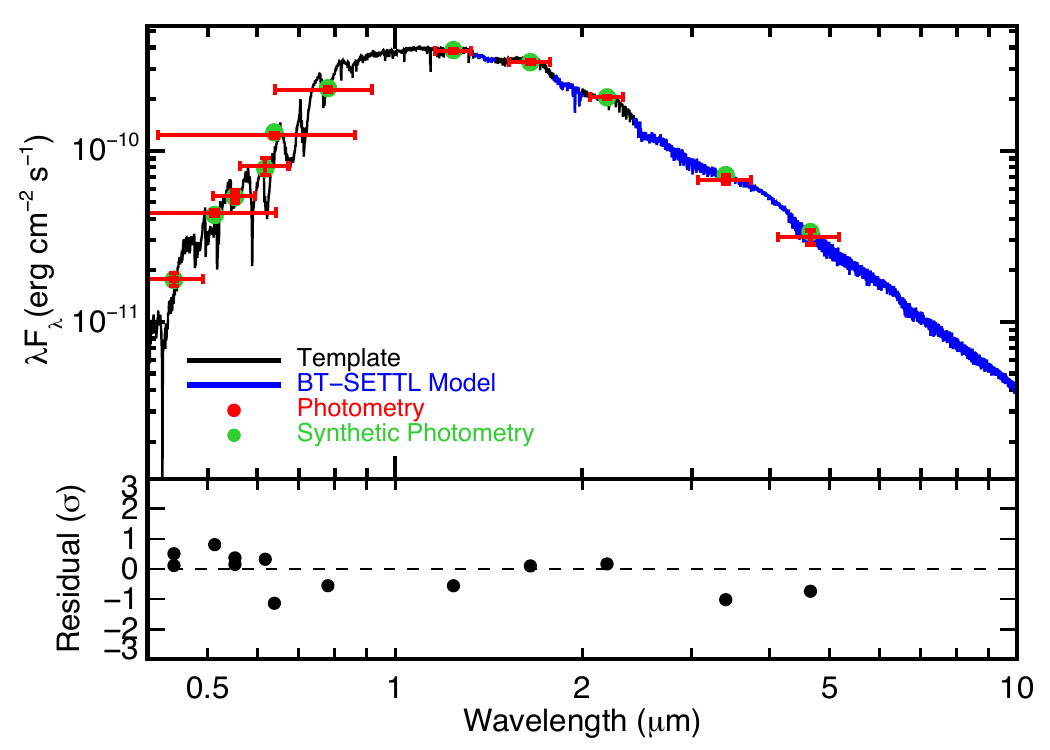} \caption{ SED of \starname\ (red points) compared to the best-fit template spectra (black) and BT-SETTL model (blue). Horizontal errors on the observed photometry approximate the filter width. The synthetic photometry used to compare to the observed photometry is shown as green points. The bottom panel shows the residual in units of standard deviations.  \label{fig:sed}}
\end{figure}

We fit the SED of \starname\ following the method in \cite{Mann2016b}, but (as above) ignoring the effects of extinction and using templates instead of a flux-calibrated spectrum. To summarize, we compared photometry summarized in Table~\ref{tab:prop} to a grid of flux-calibrated templates from \citet{Gaidos2014} and \citet{Mann2013c}. The templates spanned 0.4--2.4\um, and we used PHOENIX BT-SETTL atmosphere models \citep{Allard2013} beyond these limits or in regions of high telluric contamination (see Figure~\ref{fig:sed}). We also tested SPHINX model spectra \citep{Iyer2023}, which did a better job reproducing the template spectra but resulted in a negligible change to the final \fbol\ and \teff. The result was an absolutely-calibrated spectrum, which we used to compute \fbol\ by integrating the spectrum with wavelength. We turned this into $L_*$ using the \gaia\ DR3 parallax. We estimated \teff\ from the BT-SETTL model fit against the template and photometry as part of the comparison (the model selection is a free parameter). Finally, we estimated $R_*$ from these parameters using the Stefan-Boltzmann relation. 

The final error analysis accounted for errors in template choice, systematics in estimating \teff\ of M dwarfs models \citep{Mann2013c}, shape errors in the templates, as well as errors in the parallax and observed photometry. The final fit yielded \teff=3326$\pm$66\,K, \fbol$=(4.6\pm0.1)\times10^{-10}$\,erg\,cm$^{-2}$\,s$^{-1}$, $L_*=0.0201\pm0.0004L_\odot$, and $R_*=0.435\pm0.019R_\odot$. The final radius is consistent with the value estimated from the $M_K-R_*$ relation in Section~\ref{sec:emp}. These parameters are also consistent with those in the \tess\ Input Catalog \citep[TIC;][]{Stassun2019}. This is expected, as the methods used for the TIC M dwarfs are similar to those employed here \citep{Muirhead:2018}. Our adopted parameters are listed in Table~\ref{tab:prop}. 
 
\subsection{Stellar Inclination}\label{sec:stellar_inc}

It is possible to test whether the stellar spin and planetary orbit are in alignment by converting estimates of \vsini, \prot, and $R_*$ into an estimate of the stellar inclination ($i_*$). A simplified version of this conversion can be done by estimating the equatorial velocity ($V$) in \vsini\ using $V=2\pi R_*/P_{\rm{rot}}$. In practice, this requires additional statistical corrections, including the fact that we can only measure alignment projected onto the sky. To this end, we followed the formalism from \citet{2020AJ....159...81M}. The resulting stellar inclination was consistent with alignment with the planet, yielding a limit of $>75^{\circ}$ at 95\% and $>82^{\circ}$ at 68\%.

\begin{deluxetable}{lrr}
\tabletypesize{\scriptsize}
\tablewidth{0pt}
\tablecaption{Properties of the host star TOI 1224 \label{tab:prop}}
\tablehead{\colhead{Parameter} & \colhead{Value} & \colhead{Source} }
\startdata
\multicolumn{3}{c}{Identifiers}\\
\hline 
Gaia & 4620009665047355520 & \emph{Gaia} DR3 \\
TIC &  299798795 & \citet{stassun2018tess} \\
2MASS & J02284641-8053571 & \citet{cutri20032mass}\\
ALLWISE & J022847.15-805356.9 & ALLWISE\\ 
\hline
\multicolumn{3}{c}{Astrometry}\\
\hline
$\alpha$  & 37.19797662402810 &\emph{Gaia} DR3 \\
$\delta$ & -80.89910153220920 & \emph{Gaia} DR3\\ 
$\mu_\alpha$ (mas\,yr$^{-1}$) &  159.965$\pm$ 0.022 &\emph{Gaia} DR3\\
$\mu_\delta$  (mas\,yr$^{-1}$) & 27.171$\pm$ 0.022 & \emph{Gaia} DR3\\
$\pi$ (mas) &  26.827$\pm$ 0.018 & \emph{Gaia} DR3\\ 
\hline
\multicolumn{3}{c}{Photometry}\\
\hline
$G_{Gaia}$ (mag) & 12.768$\pm$ 0.0029  & \emph{Gaia} DR3\\
$BP_{Gaia}$ (mag) & 14.176$\pm$ 0.006 & \emph{Gaia} DR3\\ 
$RP_{Gaia}$ (mag) & 11.601$\pm$ 0.005 & \emph{Gaia} DR3\\
$B$ (mag) & 15.513$\pm$0.09 & UCAC4 \\ 
$V$ (mag) & 13.950$\pm$0.02 & UCAC4 \\ 
$R$ (mag) & 13.604$\pm$0.05 & UCAC4 \\ 
$g$' (mag) & 14.697$\pm$0.04 & UCAC4 \\ 
$r$' (mag) & 13.336$\pm$0.03 & UCAC4 \\ 
$i$' (mag) & 12.014$\pm$0.07 & UCAC4 \\ 
$J$ (mag) & 10.018$\pm$0.023 & 2MASS \\ 
$H$ (mag)  &  9.405$\pm$0.024 & 2MASS \\	 
$K_S$ (mag) & 9.134$\pm$0.019 & 2MASS \\ 
$W1$ (mag) & 9.036$\pm$0.022& ALLWISE \\
$W2$ (mag) &  8.890$\pm$0.019 & ALLWISE \\
$W3$ (mag) & 8.758$\pm$0.023 & ALLWISE \\
\hline
\multicolumn{3}{c}{Kinematics \& Position}\\
\hline
RV$_{\rm{Bary}}$ (km\,s$^{-1}$) & 14.63$\pm$1.92 & \emph{Gaia} DR3 \\
$U$ (km\,s$^{-1}$) & $-16.16\pm0.74$ & This work\\  
$V$ (km\,s$^{-1}$) & $-27.81\pm1.38$ & This work\\
$W$ (km\,s$^{-1}$) & $-1.59\pm1.11$ & This work\\
$X$ (pc) & $14.437\pm0.010$ & This work\\
$Y$ (pc) & $-26.759\pm0.018$ & This work\\
$Z$ (pc) & $-21.565\pm0.015$ & This work\\
\hline
\multicolumn{3}{c}{Physical Properties}\\
\hline
$P_{\rm{rot}}$ (days) & $1.230$ & This work  \\
\vsini (km\,s$^{-1}$) & $22.1\pm1.2$ & This work \\ 
$i_*$ ($^\circ$) & $>82$ & This work\\
\fbol\ ($\times 10^{-8} $\,erg\,cm$^{-2}$\,s$^{-1}$) & 0.046 $\pm$ 0.001 & This work\\ 
\teff\ (K) & 3326 $\pm$ 66 & This work \\
$M_*$ $(M_\odot)$ & 0.400 $\pm$0.010 & This work \\
$R_*$ $(R_\odot)$ & $0.404\pm0.012$ & This work\\
$L_*$ ($L_\odot$)  &  $(2.01\pm0.04) \times10^{-2}$  & This work\\
$\rho_*$ $(\rho_\odot)$ & $4.86 \pm 0.66$ & This work \\
Age (Myr) &  210$\pm$27  & This work
\enddata
\end{deluxetable}

\section{Transit Analysis}\label{sec:planet}

\subsection{Identification of the transit signals} \label{sec:identification}

The inner planet, TOI-1224.01, was first detected in a joint transit search of TESS sectors 1 and 13 as part of the SPOC search using an adaptive, wavelet-based matched filter \citep{Jenkins2002, Jenkins:2010qy, Jenkins2020}. The candidate passed all performed diagnostic tests \citep{Twicken2018} and was fitted with an initial limb-darkened transit model \citep{Li2019}. The difference image centroid test located the host star within 3.0$\pm$2.9 arcsec of the transit source, which was further constrained to 1.3$\pm$2.6 arcsec in a search of sectors 1-39. The TESS Science Office reviewed the diagnostic test results and issued an alert for this planet candidate as a TESS object of interest (TOI) on 2019 August 26 \citep{2021ApJS..254...39G}. 

This particular case posed challenges for the SPOC due to stellar oscillations, as TPS is specifically designed to handle colored broad-band noise. To address this, we refrained from identifying and removing sinusoidal harmonics, a technique applied in \kepler\, missions, as it tended to inadvertently eliminate energy from transit signatures, particularly those with shorter periods. Notably, the first transit of TOI-1224 c identified by the SPOC in Sector 68 is 102 orbital periods after the epoch for this planet in the manuscript. However, this discovery was after the identification of planet c through a custom search, as elaborated below.

We searched for additional planets using the Notch and LoCoR pipelines (\texttt{N\&L}). \texttt{N\&L} is described in further detail in \citet{Rizzuto2017}\footnote{\faGithub \url{https://github.com/arizzuto/Notch_and_LOCoR}} and has been used widely to search for young planets in light curves from \kepler\ \citep[e.g.,][]{Barber2022}, \ktwo\ \citep[e.g.,][]{Mann2017a}, and \tess\ \citep[e.g.][]{THYMEII}. 

In addition to the original planet at 4.178\,days, we identified an additional transit-like signal at 17.945\,days with high SNR (72) when including all \tess\ data. The only other signals that passed our SNR threshold ($>$13) were rejected as clear aliases of the other two planets. 

\subsection{TOI-1224 b planet parameters}

We used \texttt{MISTTBORN} (MCMC Interface for Synthesis of Transits, Tomography, Binaries, and Others of a Relevant Nature) \footnote{\faGithub  \url{https://github.com/captain-exoplanet/misttborn}} to fit the transit photometry for planet b. \texttt{MISTTBORN} was first detailed in \citet{Mann2016a} with significant expansion detailed in \citet{MISTTBORN}. It uses \texttt{BATMAN} \citep{Kreidberg2015} for generating the model light curves, \texttt{emcee} \citep{Foreman-Mackey2013} to explore the transit parameter space using an affine-invariant Markov chain Monte Carlo (MCMC) algorithm, and \texttt{celerite} \citep{celerite} to model the stellar variability with a Gaussian process (GP). 

We initially fit both planets using \texttt{MISTTBORN}, but found that planet c undergoes transit timing variations (TTVs). \texttt{MISTTBORN} is not setup for TTVs, so we opted to run the c planet using \texttt{juliet} \citep{espinoza2019juliet} (see Section~\ref{sec:c}). We also tested running planet b with \texttt{juliet} for TTVs as a check and found the results were consistent and no significant TTV was detected.

\begin{figure}[ht]
    \centering 
    \includegraphics[width=0.45\textwidth]
    {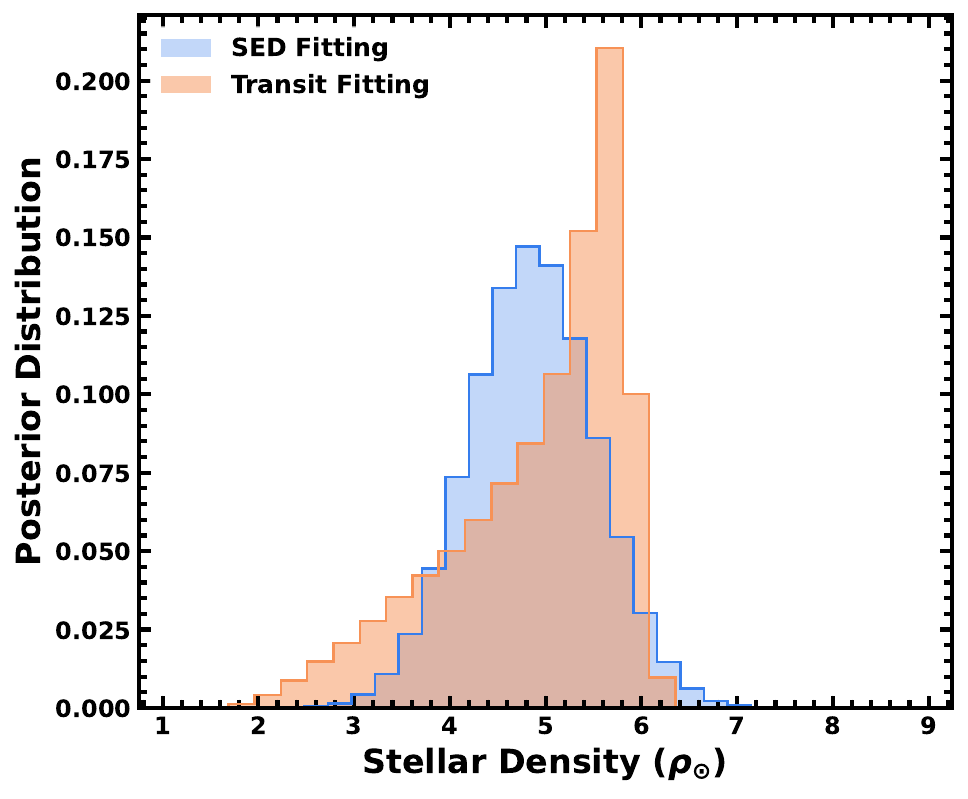} 
    \caption{Stellar density from SED fitting (blue) compared to the stellar density from the transit fit (orange) assuming the eccentricity is 0. This is consistent with a low or zero eccentricity for TOI-1224 b, which is common for (older) multi-planet systems \citep{vanEylen_eccentricity2019}.
     \label{fig:stellar_density}}
\end{figure}

We fit 14 parameters in total. The first four were the regular transit parameters: time of inferior conjunction ($T_{0}$), the orbital period of the planet ($P$), planet-to-star radius ratio ($R_p/R_\star$), and impact parameter ($b$). The fifth parameter was stellar density ($\rho_\star$). Eccentricity was assumed to be zero, but we test this assumption by comparing the density from Section~\ref{sec:stellar_params} to that from the transit (Figure ~\ref{fig:stellar_density}). 

For the limb-darkening relation, we assumed quadratic ($q_1$, $q_2$) following the triangular sampling prescription of \citet{Kipping2013}. We included data from three unique bands: \textit{TESS}, $g_{p}$', and $z_s$, requiring six limb-darkening parameters in total (see Table ~\ref{tab:limb_darkening_priors}). 

To model stellar variations, \texttt{misttborn} includes a Gaussian Process (GP) regression module, utilizing the \texttt{celerite} code \cite{foreman2017celerite}. We initially used a mixture of two stochastically driven damped simple harmonic oscillators (SHOs) at the primary period ($P_{GP}$) and secondary period ($0.5P_{GP}$). However, our analysis revealed that the second SHO was not necessary, prompting us to re-run the model with a single SHO. The GP model consists of three parameters: the period of the GP ($\ln{P_{GP}}$), the amplitude of the GP ($\ln{\rm{Amp}}$), and the decay timescale for the variability (quality factor, $\ln{Q}$). All the GP parameters were explored in logarithmic space. 

We applied Gaussian priors on the limb-darkening coefficients based on the values from the \texttt{LDTK} toolkit \citep{2015MNRAS.453.3821P}, with errors accounting for errors in stellar parameters and the difference between models used (which differ by 0.04-0.08). A summary of the priors on limb darkening coefficients can be seen in Table ~\ref{tab:limb_darkening_priors}. We also applied a Gaussian prior on the stellar density, based on our stellar parameters in Section~\ref{sec:stellar_params}. All other parameters were sampled uniformly with physically motivated boundaries (e.g., $|b|<1$, $0<R_P/R_*<1$, and $\rho_*>0$). 

\begin{figure*}[ht]
    \centering 
    \includegraphics[width=0.49\textwidth]
    {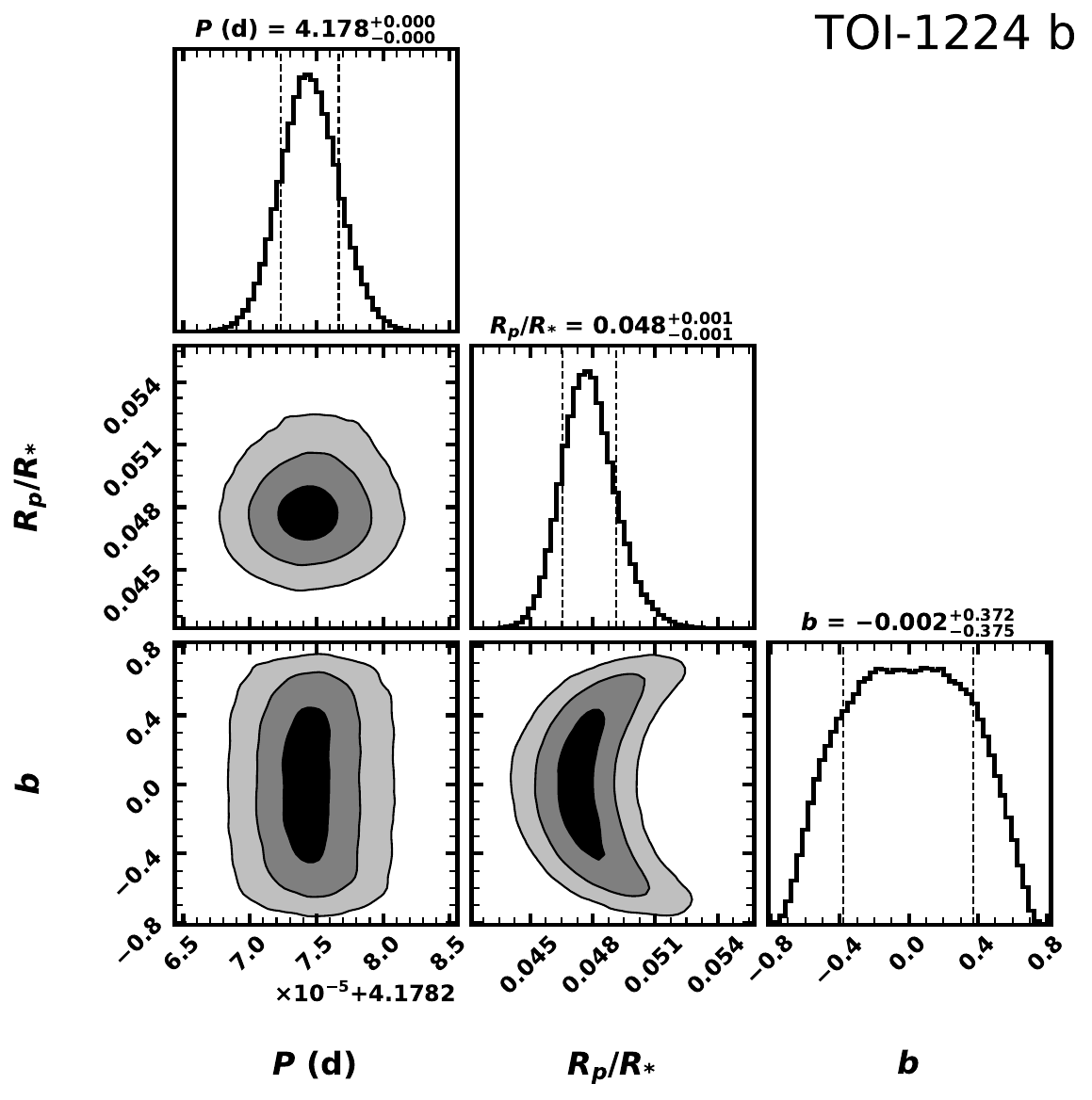} 
\includegraphics[width=0.49\textwidth]
    {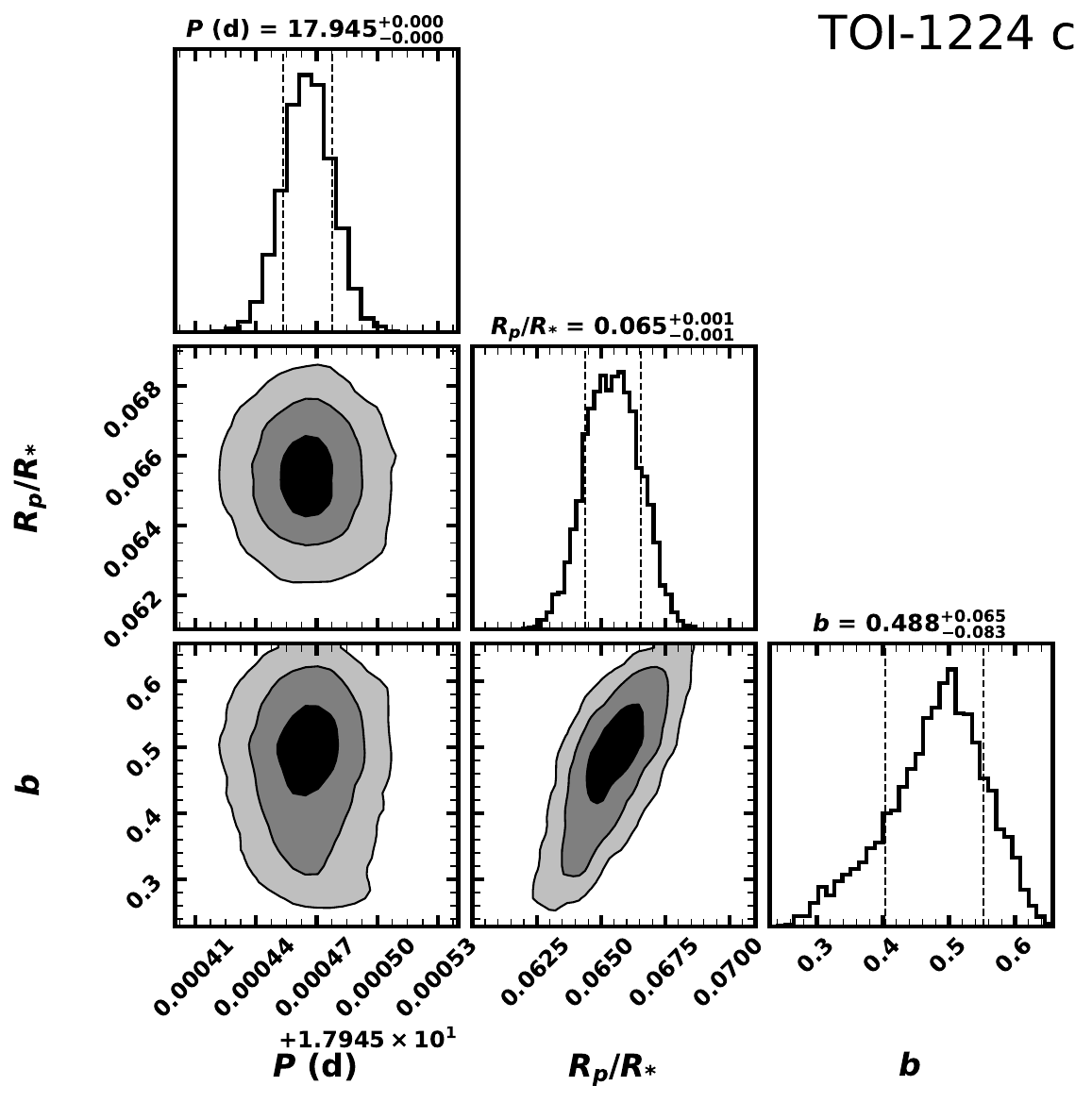} 
    \caption{Posteriors from the MCMC fit for the parameters, period ($P$), planet-to-star radius ratio ($R_P/R_*$), and impact parameter($b$) for planet b (left) and planet c (right). Planet b was fit using \texttt{MISTTBORN} and planet c was fit using \texttt{juliet} to account for the TTVs. In the histogram, the dashed lines indicate the 16\% and 84\% percentiles. Figure made with \texttt{corner.py} \citep{foreman2016corner}. 
    \label{fig:transit_corner}}
\end{figure*}

We ran the MCMC using 100 walkers for 150,000 steps including a burn-in of 20,000 steps. This run was more than 50 times the autocorrelation time for all parameters, indicating it was more than sufficient for convergence. All output parameters from the \texttt{MISTTBORN} analysis are listed in Table~\ref{tab:transitfit}, with a subset of the planetary parameter correlations in Figure~\ref{fig:transit_corner}.

\begin{figure*}[ht]
    \centering
    \includegraphics[width=0.97\textwidth]{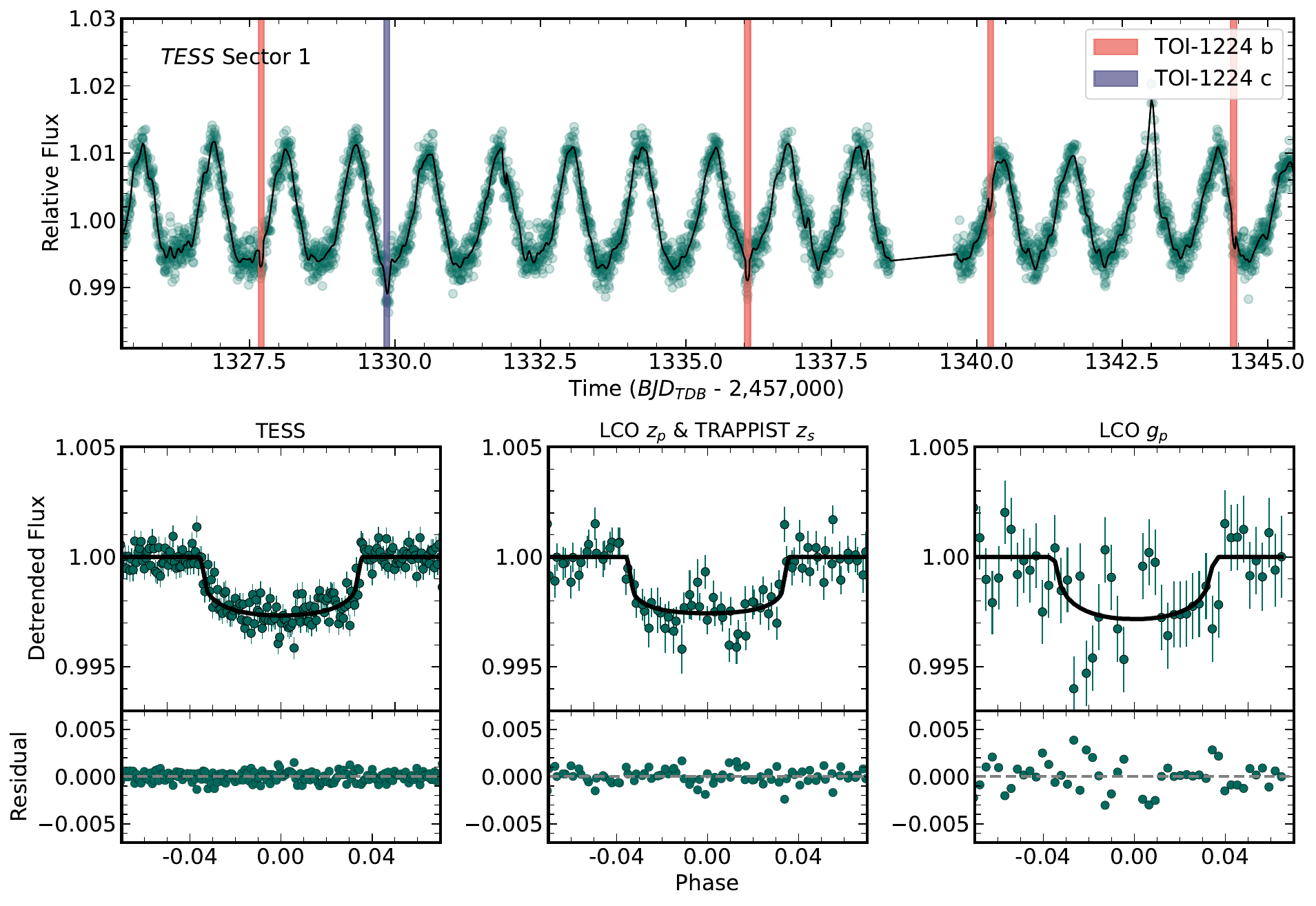} \caption{\textit{Top: } \tess\, Sector 1 light curve (green points) is binned to 5 minute intervals and is overplotted with a Gaussian process for the stellar variability (black line). Only the transits of planet b is modeled here. The transit times of the two planets are marked in purple for TOI-1224 b and pink for TOI-1224 c. \textit{Bottom: } Phase folded light curve of planet b after the best fit stellar variability model has been removed taken in filters \tess\,, $z$, and $g_{p}$. Data for \tess\, and $z$ is binned into 10 and 5 minute bins, with the best-fit transit model illustrated by a black line. } 
    \label{fig:planet_b_lc}
\end{figure*}

\begin{deluxetable*}{lcr}
\tablewidth{0pt}
\tablecaption{Parameters of TOI-1224 b \label{tab:transitfit}
}
\tablehead{
\colhead{Description} & \colhead{Parameter} & \colhead{b} \\[-0.3cm] }
\startdata
\multicolumn{3}{c}{Fit Parameters }\\
\hline
First mid-transit midpoint & $T_0$ (BJD) & $2458327.70236^{+0.00055}_{-0.00054}$ \\ 
Orbital Period & $P$ (days) & $4.1782745^{+2.2\times10^{-6}}_{-2.1\times10^{-6}}$ \\ 
Planet-to-star radius ratio  & $R_P/R_{\star}$ & $0.0478^{+0.0013}_{-0.0012}$ \\ 
Impact Parameter & $b$ & $0.27^{+0.22}_{-0.18}$ \\ 
Stellar density & $\rho_{\star}$ ($\rho_{\odot}$) & $5.19^{+0.55}_{-1.0}$ \\ 
\textit{TESS} Limb darkening coefficient & $q_{1}$ & $0.416^{+0.094}_{-0.089}$ \\ 
\textit{TESS} Limb darkening coefficient & $q_{2}$ & $0.276^{+0.071}_{-0.072}$ \\ 
\textit{LCO$z_{s}$} Limb darkening coefficient & $q_{1}$ & $0.320 \pm 0.110$ \\ 
\textit{LCO$z_{s}$} Limb darkening coefficient & $q_{2}$ & $0.202^{+0.077}_{-0.076}$ \\ 
\textit{LCO$g_{p}$} Limb darkening coefficient & $q_{1}$ & $0.750 \pm 0.120$ \\ 
\textit{LCO$g_{p}$} Limb darkening coefficient & $q_{2}$ & $0.297 \pm 0.054$ \\ 
\hline
\multicolumn{3}{c}{GP Parameters }\\
\hline
Log period & $\log(P_{GP})$ & $1.9^{+0.54}_{-0.45}$ \\ 
Log variability amplitude  &  $\log(Amp)$ & $-9.841^{+0.053}_{-0.052}$ \\ 
Quality Factor of main period oscillator &  $\log(Q_{1})$ & $0.5022^{+0.0033}_{-0.0015}$ \\ 
\hline
\multicolumn{3}{c}{Derived Parameters }\\
\hline
Ratio of semi-major axis to stellar radius & $a/R_{\star}$ & $18.9^{+0.64}_{-2.0}$ \\ 
Inclination & $i$ ($^{\circ}$) & $89.19^{+0.56}_{-0.8}$ \\ 
Transit depth & $\delta$ (\%) & $0.228^{+0.013}_{-0.011}$ \\ 
Planet radius & $R_P$ ($R_\oplus$) & $2.104^{+0.094}_{-0.091}$\\
Semi-major axis & $a$ (AU) & $0.0355^{+0.0017}_{-0.0034}$ \\ 
Equilibrium Temperature \tablenotemark{a} & $T_{\mathrm{eq}}$ (K) & $541.0^{+26.0}_{-14.0}$ \\ 
\hline 
\enddata
\tablenotetext{a}{Assumes an albedo of 0}
\end{deluxetable*}


\begin{deluxetable} {lcccccr}
\tabletypesize{\scriptsize}
\tablecaption{Priors on Limb-darkening Coefficients \label{tab:limb_darkening_priors}}
\tablecolumns{3}
\tablewidth{0pt}
\tablehead{
\colhead{Filter} & 
\colhead{$g_{1}$} &  
\colhead{$g_{2}$}}
\startdata
TESS & 0.272$\pm$0.08 & 0.317$\pm$0.04\\ 
$z_{s}$ & 0.232$\pm$0.08 & 0.330$\pm$0.04\\ 
$g_{p}$ & 0.533$\pm$0.08 & 0.299$\pm$0.04\\ 
\enddata
\tablecomments{Limb-darkening priors are provided as the traditional linear and quadratic terms but were fit using triangular sampling terms.}
\end{deluxetable}

\subsection{TOI-1224 c planet parameters and TTV} \label{sec:c}

To measure the transit parameters and account for TTVs, we used \texttt{juliet} (Joint Analysis of Exoplanetary Transits \& RVs) on the \tess\, and ground-based light curves. \texttt{juliet} \citep{espinoza2019juliet}  \footnote{\faGithub \url{https://github.com/nespinoza/juliet/}} is a Python transit fitting package that utilizes the python package, \texttt{batman} \citep{Kreidberg2015} to model the transits, \texttt{celerite} \citep{celerite} to model the stellar variability, and \texttt{dynesty}  \citep{sergey_koposov_2023_7600689, speagle2020dynesty} \footnote{\faGithub \url{https://github.com/joshspeagle/dynesty}} to perform nested sampling of the posteriors.

\subsubsection{\tess\, light curves}

\texttt{juliet} can simultaneously model the stellar variability, the transit, and the transit timing variations. We used the \tess\, light curves after the removal of flares (Section ~\ref{sec:tess}). We then used a Matern Gaussian Process (GP) kernel implemented in \texttt{celerite} \citep{celerite} to model the stellar varability. This feature as available in the \texttt{juliet} code, placing broad priors on the two GP parameters, amplitude of the GP ($\sigma_{GP}$) and length-scale of the GP ($\rho_{GP}$). The GP parameters were explored in logarithmic space.

In addition to the GP parameters, there are three required instrumental parameters: m$_{dilution}$ (dilution factor), m$_{flux}$ (offset relative flux), and $\sigma$ (jitter term to account for additional systematic). Since no external contaminating sources were present, the dilution factor was fixed at 1. 

To refine the Gaussian Process model, we identified and removed $\sim$ 130 data points corresponding to a flare occurring immediately after the ingress of transit number 99 ($T_{99}$), which was missed by \texttt{stella}. This particular subject falls within the time range of 2460106.345 and 2460106.386 BJD$_{TDB}$ and was impacting the GP. 

\subsubsection{Ground-based data}

There were four total TOI-1224 c transits taken with ground-based data: two taken with ASTEP in filter $R$, one taken by LCO in filter $z_{s}$ and another transit taken by LCO in filter $g_{p}$ (refer to Table ~\ref{tab:obslog} for details). All the ground-based data were normalized and detrended. Each dataset included the three essential instrumental parameters ($m_{flux}$, $m_dilution$, and $\sigma$). The dilution factor for each dataset was set to 1.

\subsubsection{TTV Fitting}
\begin{figure*}[ht]
    \centering
    \includegraphics[width=0.90\textwidth]{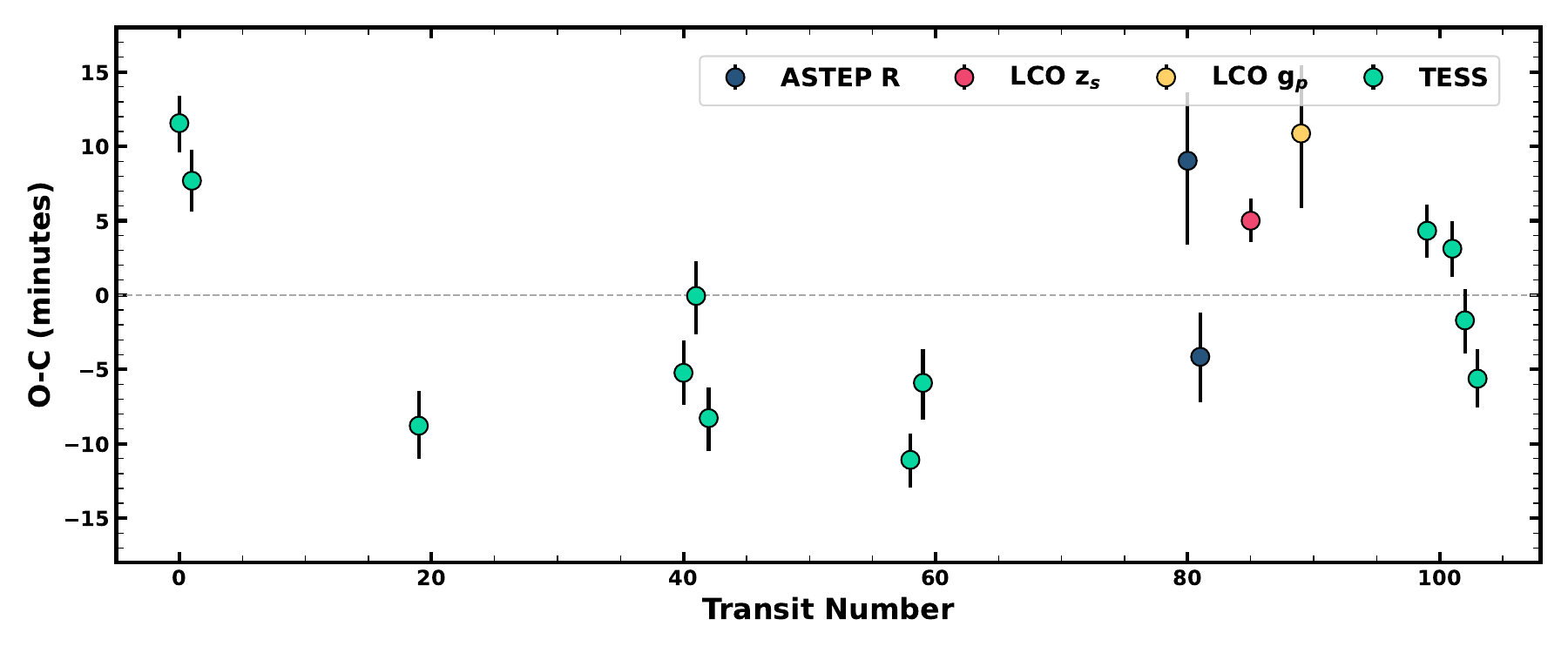} \caption{Observed minus calculated (O-C) mid transit times as a function of time for TOI-1224 c in minutes taken with \tess\, (green), ASTEP $R$ (purple), LCO $z_{s}$ (red), and LCO $g_{p}$(yellow). The gray dashed horizontal line represents a linear ephemeris. 
    \label{fig:oc_plot}}
\end{figure*} 

Following \cite{espinoza2018efficient}, instead of fitting for the planet-to-star radius ratio ($p=R_{p}/R_{*}$) and impact parameter ($b$), we fit for the parameters for $r_{1}$ and $r_{2}$.  We placed a uniform prior on both these parameters. We adapted a quadratic limb-darkening law, where we place a uniform prior on both coefficients \citep[$q_{1}$ and $q_{2}$; ][]{Kipping2013}. We set the eccentricity ($e$) and the argument of periastron ($\omega$) to 0 and 90 \degree. We placed a Gaussian prior on the the stellar density ($\rho_{\odot}$). To model the transit time ($T_{n}$), we adopted Gaussian priors with a width of 1 hour for each of the transits. With 16 transits total (12 \tess\, and 4 ground-based), a dedicated parameter ($T_{n}$) was assigned, resulting in a total of 16 $T$ values. 

\begin{figure}[ht]
    \centering
    \includegraphics[width=0.50\textwidth]{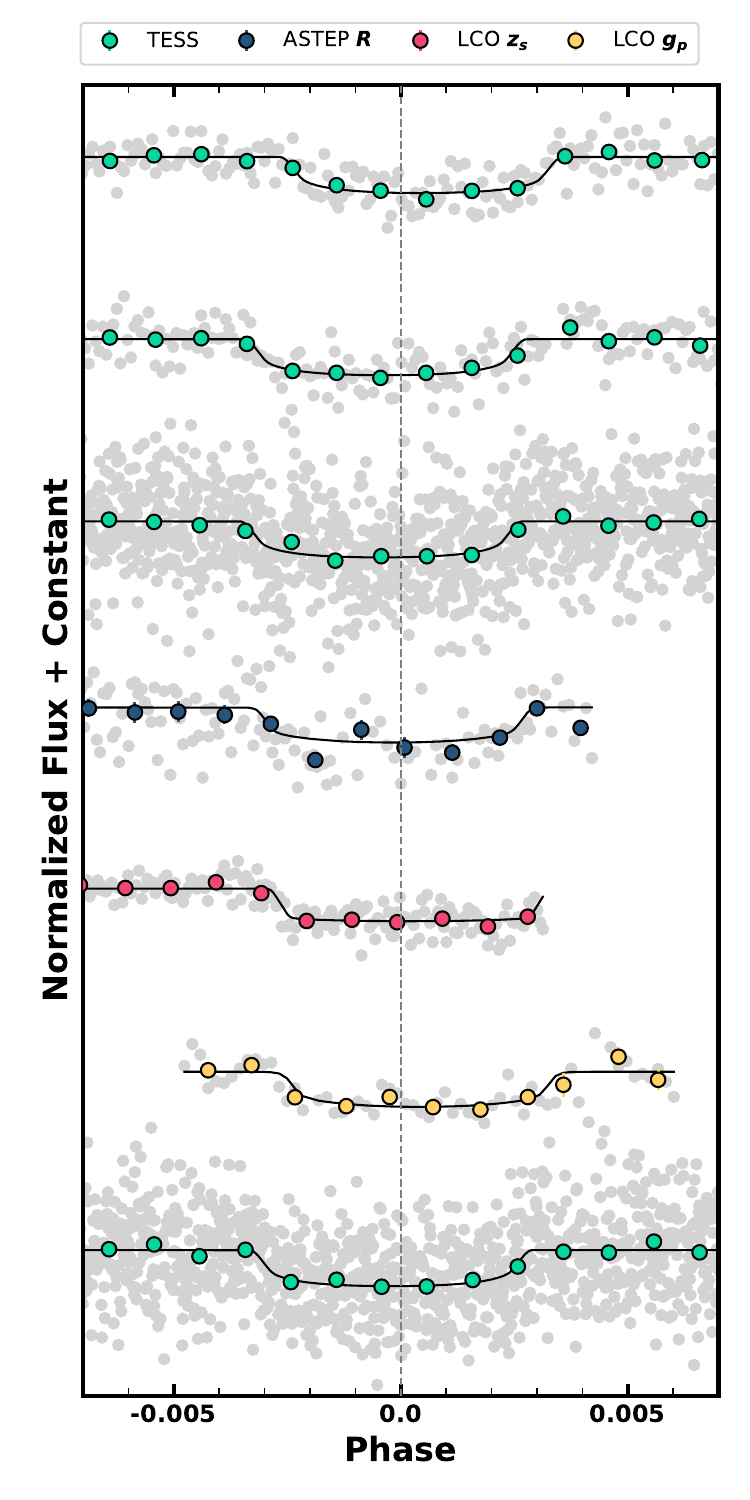} \caption{Phase-folded light curve of TOI-1224 c taken with \tess\, (green), ASTEP $R$, (blue), LCO $z_{s}$ (red) and, LCO $g_{p}$ (yellow). Transit numbers 0, 19, 58, 81, 85, 89, 103 are arranged from top to bottom and are offset for clarity. The best-fit transit model is overlaid as a black line. Circular points is data binned into 1.5 minute intervals.
    \label{fig:individual_lc}}
\end{figure}

The overall model comprised a total of 35 parameters. The log evidence ln($Z$) for this fit was 6.35$\pm$0.19. The transit epoch for planet c is 2458329.8597 $\pm$ 0.0008  $BJD_{TDB}$ with a period of 17.945466 $\pm$0.000012 days. The planet c parameters and resulting observed and calculated time are presented in Table ~\ref{tab:ttv}. Visualizing the amplitude of the Transit Timing Variations (TTVs), Figure~\ref{fig:oc_plot} offers a graphical representation, while Figure~\ref{fig:individual_lc} showcases a subset of individual transits. The transit timing variations of TOI-1224 c deviates from a linear ephemeris by $\sim$10 minutes. With this detection, this system joins the list of young planetary systems exhibiting such phenomena \citep[e.g., TOI-1227; ][]{almenara2024evidence}, and heightens the prospects for further investigation.

\begin{deluxetable*}{lclrr} 
\tabletypesize{\footnotesize} 
\tablecaption{Prior and Posterior Distributions for TTV Extraction using \texttt{juliet} for TOI-1224 c \label{tab:ttv}}
\tablewidth{0pt}
\tablehead{
\colhead{Description } &
\colhead{Parameter} & 
\colhead{Prior\tablenotemark{a}}  &
\colhead{Posterior} & 
\colhead{O-C (min)}} 
\startdata
    \multicolumn{4}{c}{\tess\, TTV Fitting  }\\ 
    \hline 
Limb darkening & $q_{1}$ & $\mathcal{U}$ (0.0, 1.0) & $0.61^{+0.24}_{-0.25}$ & \\
Limb darkening & $q_{2}$ & $\mathcal{U}$ (0.0, 1.0) & $0.14^{+0.18}_{-0.10}$ & \\
Offset relative flux & $m_{flux}$ & $\mathcal{N}$ (0.0, 0.1) & $-0.0001^{+0.0}_{-0.0}$ & \\
Jitter & $\sigma$ & $\mathcal{J}$ (0.1, 1000.0) & $548.52^{+31.43}_{-32.41}$ & \\
Amplitude of GP & GP$_{\sigma}$ &  $\mathcal{J}$ (0.000001, 1000000) &  
0.0046$\pm$0.0001 \\ 
Time of the GP & GP$_{\rho}$ &  $\mathcal{J}$ (0.000001, 1000000) &  0.1511$^{+0.0036}_{-0.0035}$\\ 
Time of Transit Center & $T_{0}$ & $\mathcal{N}$ (2458329.8653, 0.0417) & $2458329.8678^{+0.0015}_{-0.0016}$ & $11.56^{+1.83}_{-1.98}$  \\ 
& $T_{1}$ & $\mathcal{N}$ (2458347.81055, 0.0417) & $2458347.8105^{+0.0017}_{-0.0017}$ &  $7.69^{+2.09}_{-2.06}$ \\ 
& $T_{19}$ & $\mathcal{N}$ (2458670.8250, 0.0417) & $2458670.8174^{+0.0018}_{-0.0017}$ & $-8.78^{+2.31}_{-2.21}$  \\ 
& $T_{40}$ & $\mathcal{N}$ (2459047.6751, 0.0417) & $2459047.6747^{+0.0015}_{-0.0016}$ & $-5.23^{+2.18}_{-2.15}$  \\
& $T_{41}$ & $\mathcal{N}$ (2459065.6203, 0.0417) & $2459065.6238^{+0.0016}_{-0.0018}$ & $-0.06^{+2.33}_{-2.57}$  \\ 
& $T_{42}$ & $\mathcal{N}$ (2459083.56560, 0.0417) & $2459083.5636^{+0.0014}_{-0.0016}$ &  $-8.27^{+2.08}_{-2.21}$  \\ 
& $T_{58}$ & $\mathcal{N}$ (2459370.6895, 0.0417) & $2459370.689^{+0.0012}_{-0.0013}$ &  $-11.08^{+1.78}_{-1.88}$  \\ 
& $T_{59}$ & $\mathcal{N}$ (2459388.6348, 0.0417) & $2459388.6381^{+0.0016}_{-0.0018}$ & $-5.9^{+2.27}_{-2.47}$  \\ 
& $T_{99}$ & $\mathcal{N}$ (2460106.4446, 0.0417) & $2460106.4638^{+0.0011}_{-0.0011}$ & $4.33^{+1.77}_{-1.81}$  \\ 
& $T_{101}$ & $\mathcal{N}$ (2460142.3350, 0.0417) & $2460142.3539^{+0.0012}_{-0.0013}$ & $3.12^{+1.83}_{-1.87}$  \\ 
& $T_{102}$ & $\mathcal{N}$ (2460160.2803, 0.0417) & $2460160.296^{+0.0014}_{-0.0015}$ & $-1.70^{+2.09}_{-2.23}$  \\ 
& $T_{103}$ & $\mathcal{N}$ (2460178.2255, 0.0417) & $2460178.2388^{+0.0013}_{-0.0014}$ & $-5.62^{+1.98}_{-1.93}$ \\
\hline
    \multicolumn{4}{c}{ASTEP TTV Fitting  }\\
    \hline
Limb darkening & $q_{1}$ & $\mathcal{U}$ (0.0, 1.0) & $0.31^{+0.31}_{-0.21}$ \\
Limb darkening & $q_{2}$ & $\mathcal{U}$ (0.0, 1.0) & $0.41^{+0.33}_{-0.27}$ \\
Offset relative flux & $m_{flux}$ & $\mathcal{N}$ (0.0, 0.1) & $0.0005^{+0.0002}_{-0.0002}$ \\
Jitter & $\sigma$ & $\mathcal{J}$ (0.1, 1000.0) & $4.81^{+98.60}_{-4.46}$ \\
Time of Transit Center & $T_{80}$ & $\mathcal{N}$ (2459765.4849, 0.0417) & $2459765.5033^{+0.0033}_{-0.0041}$ &  $9.03^{+4.61}_{-5.61}$  \\
& $T_{81}$ & $\mathcal{N}$ (2459783.4301, 0.0417) & $2459783.4396^{+0.0021}_{-0.0022}$ & $-4.15^{+2.98}_{-3.05}$  \\ 
\hline
\multicolumn{4}{c}{LCO$z_{s}$ TTV Fitting} \\
\hline
Limb darkening & $q_{1}$ & $\mathcal{U}$ (0.0, 1.0) & $0.04^{+0.07}_{-0.03}$ \\
Limb darkening & $q_{2}$ & $\mathcal{U}$ (0.0, 1.0) & $0.32^{+0.35}_{-0.24}$ \\
Offset relative flux & $m_{flux}$ & $\mathcal{N}$ (0.0, 0.1) & $0.0003^{+0.0001}_{-0.0001}$ \\
Jitter & $\sigma$ & $\mathcal{J}$ (0.1, 1000.0) & $931.92^{+48.13}_{-76.50}$ \\
Time of Transit Center & $T_{85}$ & $\mathcal{N}$ (2459855.2111, 0.0417) & $2459855.2278^{+0.0009}_{-0.0009}$ &  $5.0^{+1.49}_{-1.45}$ \\ 
\hline
\multicolumn{4}{c}{LCO$g_{p}$ TTV Fitting}\\ 
\hline
Limb darkening & $q_{1}$ & $\mathcal{U}$ (0.0, 1.0) & $0.27^{+0.29}_{-0.19}$ \\
Limb darkening & $q_{2}$ & $\mathcal{U}$ (0.0, 1.0) & $0.46^{+0.33}_{-0.30}$  \\
Offset relative flux & $m_{flux}$ & $\mathcal{N}$ (0.0, 0.1) & $0.0005^{+0.0003}_{-0.0003}$ \\
Jitter & $\sigma$ & $\mathcal{J}$ (0.1, 1000.0) & $11.18^{+136.10}_{-10.60}$ \\
Time of Transit Center & $T_{89}$ & $\mathcal{N}$ (2459926.9921, 0.0417) & $2459927.0137^{+0.0035}_{-0.0039}$ & $10.87^{+4.61}_{-5.0}$  \\ 
\hline
     \multicolumn{4}{c}{Derived Parameters}\\ 
     \hline
Planet-to-star radius ratio \tablenotemark{b} & $R_{p}/R_{*}$ & $\mathcal{U} (0, 1)$ & $0.065\pm0.001$ \\ 
Impact Parameter \tablenotemark{b} & $b$ & $\mathcal{U} (0, 1)$ & $0.487^{+0.065}_{-0.083}$ \\ 
Eccentricity & $e$ & Fixed & 0 \\
Argument of peristron (deg) & $\omega$ & Fixed & 90 \\  
Stellar density  & $\rho_{\odot}$ & $\mathcal{N} (4.86,0.66)$ & $4.79^{+0.66}_{-0.60}$ \\ 
Planet radius & $R_{p}(\oplus$) & -- &  2.884 $\pm$ 0.098\\ 
Period (days) & $P$ & -- & $17.945466\pm+0.000012$ \\ 
Transit Epoch (BJD$_{TDB}$) & $T_{0}$ & -- & $2458329.859704^{+0.000828}_{-0.000822}$ \\ 
\enddata
\tablenotetext{a}{$\mathcal{U} (a,b)$ indicates a uniform distribution between $a$ and $b$; $\mathcal{N} (a,b)$ indicates a normal distribution with mean $a$ and a standard deviation $b$; $\mathcal{J} (a,b)$ indicates a Jeffreys prior or log-uniform distribution between $a$ and $b$.}
\tablenotetext{b}{The transit depth and impact parameters were fit using $r_{1}$ and $r_{2}$; the values here were transformed back to the $(b,p)$ plane}
\end{deluxetable*}

\subsection{False Positive Analysis} \label{sec:fp_analysis}

For our false positive analysis, we first calculate the magnitude limit ($\Delta m$) of a potential blended source (bound or background) that could reproduce the transit signal, using the source brightness constraints described by \citet{Seager2003} and \citet{Vanderburg2019}. This depends on the ingress or egress duration compared to the transit duration and reflects the true radius ratio, independent of whether there is contaminating flux: 
\begin{align}
    \Delta m \leq 2.5\log_{10}\left(\frac{T^2_{12}}{T^2_{13}\delta}\right)\notag ,
\end{align}
where $\delta$ is the transit depth, $T_{12}$ is the ingress duration and $T_{13}$ is the time between the first and third contact. We calculate $\Delta m$ for the posterior samples for our floating eccentricity transit fit and take the 99.7\% confidence limit. We find $\Delta m <1.3$ and $<1.8$ for TOI-1224\,b and TOI-1224\,c, respectively. 

\gaia\ reports a moderately high Renormalised Unit Weight Error (RUWE; 1.475) for \starname. RUWE above 1.4 is known to be indicative of binarity \citep{Wood2021,Ziegler2020}. RUWE is known to be higher for young stars \citep{Fitton2022}, but this value is still above most members of the group. \starname\ also sits high on the \gaia\ CMD (Figure~\ref{fig:gaia_cmd}), although its position is consistent with the intrinsic spread around a single-aged population ($\simeq$0.1\,mags based on our isochrone fit; Section~\ref{sec:isochronal_age}). The combination is still suggestive. To be missed in the imaging, spectroscopy, and velocity data, such a companion would either need to be faint or land behind the star near maximum separation such that most of the velocity is in the plane of the sky. The latter scenario is unlikely, but not impossible. 

In the case of a bound companion as the source of the transit signals, the resulting signal depth would be a factor of $\simeq$4 greater for planet b and a factor of $\simeq$6 greater for planet c. Assuming the companion has the same age, this would still yield a planetary radius for both signals ($\lesssim7R_\oplus$). Although such a scenario would significantly change the inferred radii of the two planets, it does not change the planetary interpretation of the signals and hence rules out any false positive scenario involving a bound companion.

\begin{figure}[ht]
    \centering
    \includegraphics[width=0.45\textwidth]{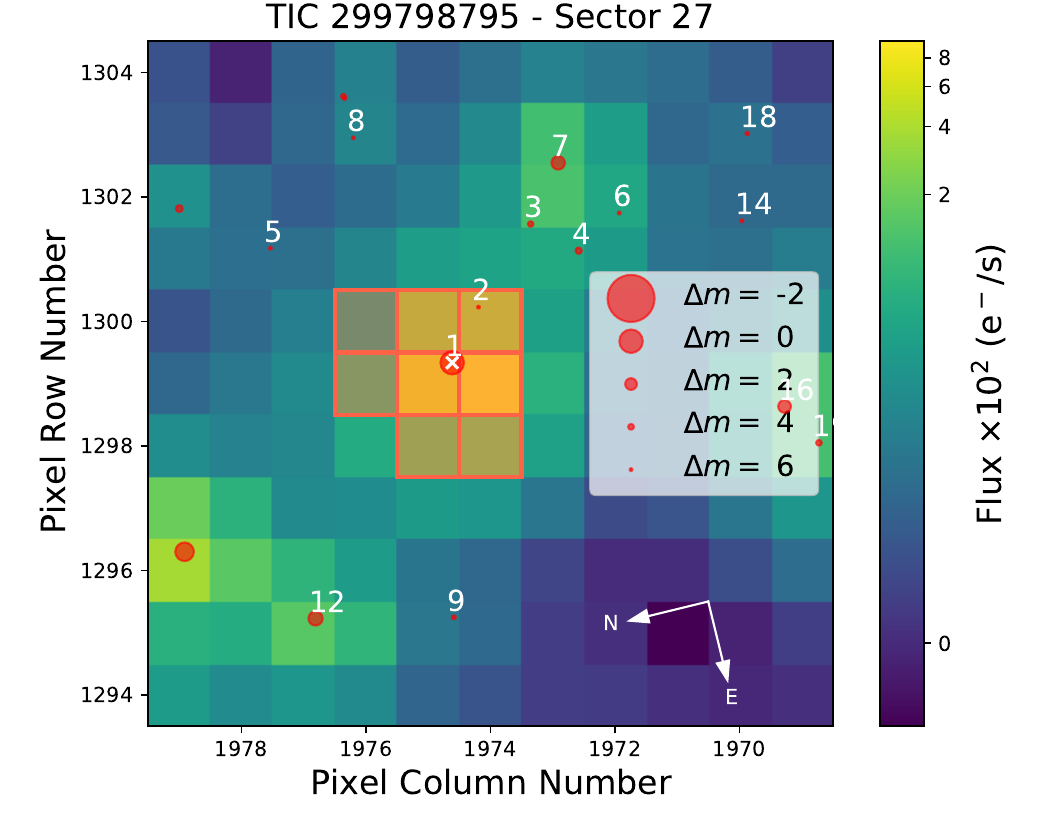} \caption{\tess\ image around \starname\ from Sector 13. The red region shows the aperture used for extracting the light curve. Points indicate other sources detected from \gaia\, scaled in size by their contrast. Other than \starname, only a single \gaia\ source is within the aperture, and it is too faint to reproduce the transits. Figure made with \texttt{tpfplotter}. \citep{Aller2020}.
    \label{fig:tpfplotter}}
\end{figure}

No \gaia\ sources within the \tess\ aperture could reproduce the observed transit shape and depth given the magnitude limits above (Figure~\ref{fig:tpfplotter}). The SOAR speckle imaging rules out any such companions down to $\simeq$0.2\arcsec. Lastly, the lack of additional lines in the CHIRON spectra rules out such bright companions assuming they are offset in velocity from \starname\ at any of the epochs. This rules out any false positive scenario involving an unassociated star.

There is also significant separate evidence that the signals are planetary. One is that the transits show consistent depths from $g'$ to $z_s$ (Figure ~\ref{fig:planet_b_lc}). As explained in \citet{Desert2015}, if the transit signals were associated with another star in the aperture, the transit depth would vary. The target shows no significant centroid offset (Section ~\ref{sec:identification}). The transit is consistently detected on the source, including in the (seeing-limited) ground-based photometry. Lastly, the baseline false-positive rate for multi-planet systems is much lower than for targets with a single detected planet \citep{LissauerAlmost2012, LissauerValidation2014}. This has been seen in \tess\ systems as well \citep{guerrero2021tess}. We conclude that both signals are unambiguously planetary in nature.

\subsection{Injection/Recovery Analysis} \label{sec:injrec}

\begin{figure}[ht]
    \centering
    \includegraphics[width=0.45\textwidth]{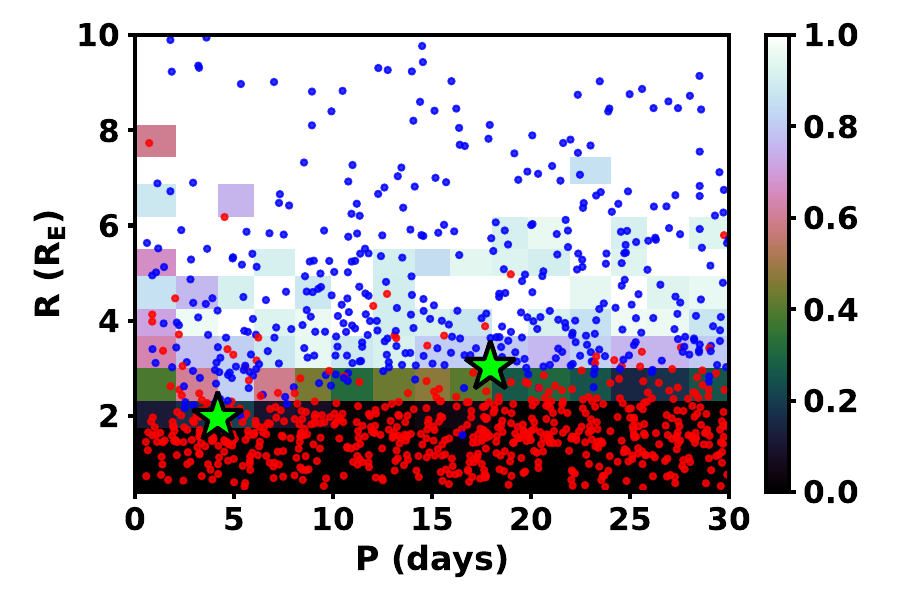} \caption{Injection/recovery test for \starname. Each point represents an injected planet, with red points indicating an injected planet we failed to recover, while  blue ones we recovered successfully. Only 20\% of injected planets are shown for clarity. The background is color-coded by the overall completeness in a given bin. Interestingly, the completeness right around the inner planet is only $\simeq$20\%
    \label{fig:injrec}}
\end{figure}

We test our sensitivity to additional planets in the system using an injection/recovery test. For this, we follow \citet{Rizzuto2017}. Specifically, we injected 5,240 simulated planets with $0.5<R_P<10$ and $0.5<P<30$. Other parameters, like the $b$ and $T_0$ are drawn from random distributions bounded by physical or data limits. For each simulated planet, we then re-run the \texttt{Notch} detection pipeline and attempt to recover the planet. The resulting completeness is shown in Figure~\ref{fig:injrec}, and suggests completeness drops off just below the smaller of the two planets.

\section{Summary \& Discussion} \label{sec:discussion}

We report the discovery and validation of two transiting sub-Neptune planets orbiting an early M-dwarf, TOI-1224. Stars around \starname\ (in space and tangential velocity) have a high fraction of consistent radial velocities (Figure~\ref{fig:gaia_cmd}), rapid rotation (Figure~\ref{fig:prot}), high lithium levels (Figure~\ref{fig:lithium}), and photometric variability \citep[Section~\ref{sec:gaiavar};][]{Barber2023} -- all of which demonstrate the group is a real young population. Following the convention from prior THYME papers \citep{THYMEV}, we name the group \association. 

To derive the age of the group, we combine measurements of lithium levels, rotation, and variability resulting in an age of \age\,$\pm$\ageerr\ Myr. We provide our list of candidate members, but with the warning that $\simeq$1/3 of the listed targets are likely non-members. 

\association\ is physically and kinematically nearby the recently reported groups, Crius 224, Theia 424 and Volans-Carina. However, these groups have assigned ages significantly younger than \association\ (80-110\,Myr versus 150-\age\,\,Myr). Further investigation is required to establish a more detailed understanding of the relationship (or lack thereof) between these populations. In future work, it would be particularly helpful to model these populations all simultaneously to separate out the real membership list. 

The youth of \starname\ is certain. It exhibits rapid rotation and H$\alpha$ emission consistent with a $<1$\,Gyr M3V \citep{Kiman2021}. However, the question of whether \starname\ is indeed a part of Crius 221, as indicated in \citet{Moranta2022}, or \association\ adds complexity to assigning a precise age. If \starname\ is a member of Crius 221, it would  account for the slightly elevated position of \starname\ in the color-magnitude diagram. If this association holds true and \starname\, is a member, it would imply that the star is significantly younger (90-100\,Myr). Fortunately, the SED-based $R_*$ works even on pre-main-sequence stars, so the age change would not impact the derived stellar (or planetary) radii, nor would it change the false-positive assessment. 

The planets TOI-1224 b and c join the growing number of planetary systems in young associations \citep[e.g.,][]{Tofflemireetal2019,THYMEIV,THYMEVII}. These planets are particularly compelling because there are only a handful of known young multi-planet systems, allowing us to test and refine planetary models within a system. In addition, there are few systems that are within the 200-400 Myr, underscoring the significance of TOI-1224 b and c in contributing to our understanding of planetary evolution during this critical epoch.

\subsection{\starname\,b and c in context}

\begin{figure}[ht]
    \centering 
    \includegraphics[width=0.5\textwidth]{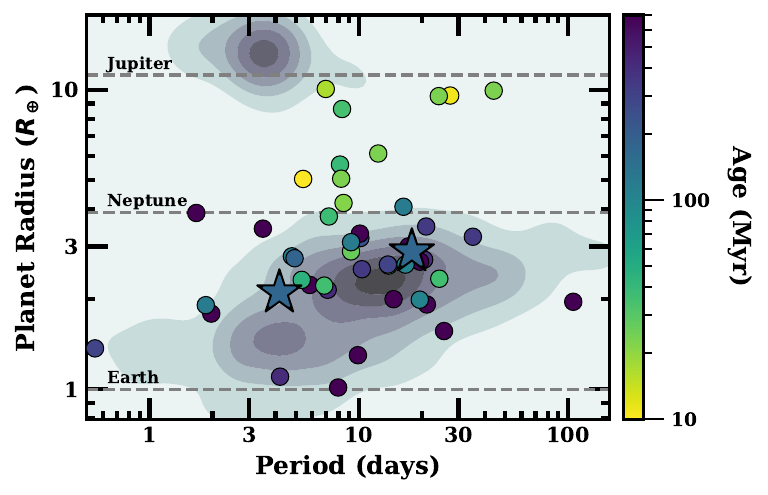}  \caption{Contour of all planets discovered by \kepler\ and \ktwo\ as a function of planet radius ($R_{\oplus}$) and orbital period (days). Young ($<700$ Myr), transiting systems are colored by their approximate age in log space, derived from their respective host cluster or association. TOI-1224 b and c are outlined as stars and fall within the distribution of mature planets. These two planets stand out as rare finds within the age range of $\sim$ 200 Myr. Planet properties from \citet{NASAexplanetarchive}. }\label{fig:young_pl}
\end{figure}

\begin{figure}[ht]
    \centering
    \includegraphics[width=0.5\textwidth]{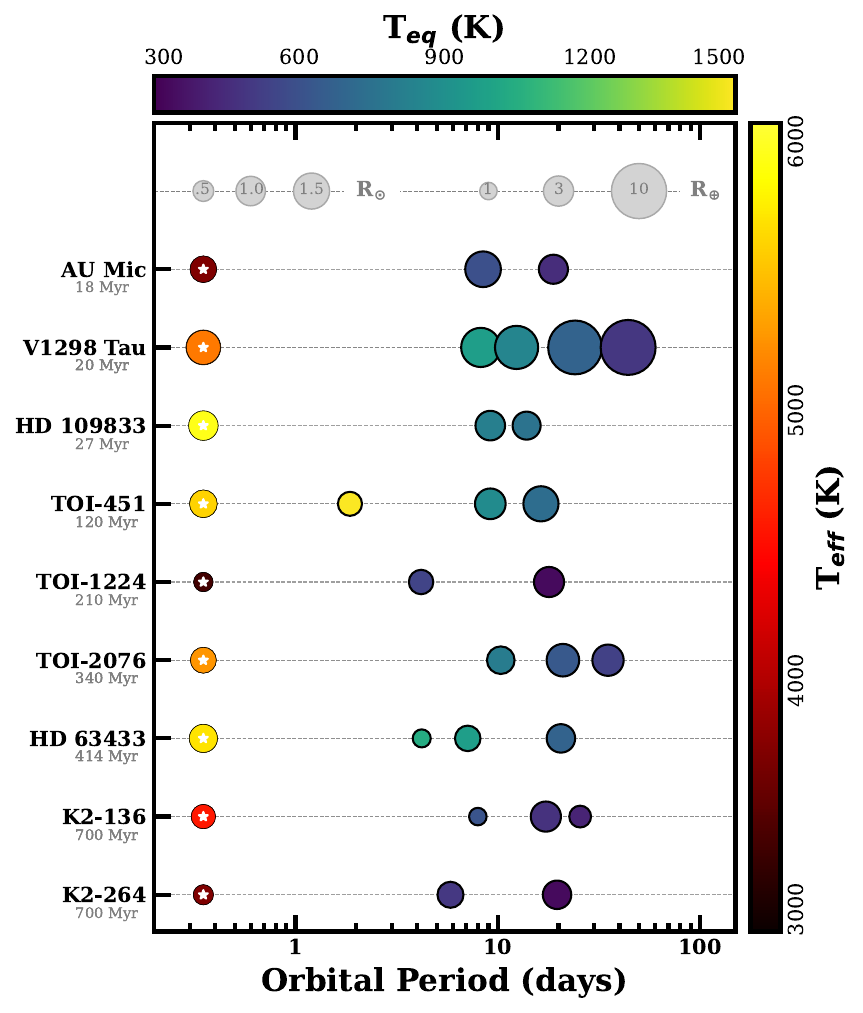}  \caption{Young ($<$ 1 Gyr), transiting, multi-planetary systems that are members of a known association or a cluster as a function of its period (days). The leftmost circle in each row represents the host star, with the size of the marker proportional to the stellar radius and colored by its  effective temperature (\teff\,). The circles to the right represent the planets in each system with the size of the marker proportional to the planetary radius and colored by its equilibrium temperature (T$_{eq}$). The systems are sorted in order of their age with the youngest systems at the top and the oldest at the bottom. Data for these planets are listed in Table~\ref{tab:young_multis}. TOI-1224 follows the intra-system uniformity pattern \citep{weiss2018california} as both planets, b and c, share comparable sizes.} \label{fig:young_multis}
\end{figure}

An increasing number of transiting planets in young associations have been discovered in the last decade (Figure~\ref{fig:young_pl}), owing mostly to the success of the \ktwo\ and \tess\ missions. The number of such multi-planet systems remains small compared to the older population, in part because \ktwo\ and \tess\ surveyed for a much shorter period than \kepler\, and the latter contained only a relatively small number of young stars and stellar associations \citep[e.g.,][]{Bouma2022}. We show all known $<750$\,Myr transiting multi-planet systems in known associations in Figure~\ref{fig:young_multis}. The list grows somewhat if we consider field stars with ages assigned from other methods, such as gyrochronology \citep{Barragan2022}, or lithium \citep{2018ApJ...855..115B}. However, many of these ages are questionable, due in part, to the recent discovery of rotation stalling \citep{Curtis_stall}, which has invalidated many earlier age relations for K dwarfs.

Old multi-planet systems are known to show a high level of intra-system uniformity: masses, radii, and orbital spacings between neighboring planets in compact multiples are more similar than expected by chance \citep{lissauer2011, 2016ApJ...822...54D, Weiss2018}. The current sample of young planets appears to display this trend. However, the sample size of young multi-planet systems is small ($\lesssim$10). Further, there are various observational biases between the two samples \citep{Rizzuto2017,Fernandes2023}. Therefore, we regard this as an interesting trend awaiting confirmation through a larger sample of systems and a more comprehensive understanding of these observational biases.

\subsection{Prospects for follow-up} 
With the launch of \textit{JWST}, we find ourselves at the forefront of atmospheric characterization in the field of exoplanets. To determine if these planets are suitable targets for observation, we employ the Transmission Spectroscopy Metric \cite[TSM;][]{kempton2018framework}. Since these young planets lack directly measured masses, we use the \texttt{forecaster} code \citep{chen2017} to estimate their masses, which assumes these young planets obey the same mass–radius relation as older stars, which may
be inaccurate. Given the radius of 2.10 and 3.00$R_{\oplus}$, the predicted masses using \texttt{forecaster} \citep{chen2017} for TOI-1224 b and c are 5.40 and 8.83$M_{\oplus}$, respectively; this resulted in TSM values of 72 and 77. It is important to note that due to their youth, these planets are typically larger and less dense compared to their mature counterparts  \citep{owen2020constraining}. Given that the true masses are likely smaller \citep[as expected for young planets;][]{Owen2018,Fernandes2022}, we re-evaluated the TSM values with a 30\% reduction in radius, resulting in revised mass estimates of 3.03 and 5.03 M$_{\oplus}$ for planets b and c. Using these values and the actual planet radii leads to an increased TSM value of 128 and 136 -- making TOI-1224 b and c promising candidates for atmospheric characterization \citep[see also][]{guillot+2022, kempton2018framework}. 

While these planets orbit a young host star, obtaining masses for these planets through RV may be challenging as RV signal from an active star can easily exceed the planetary signal by several orders of magnitude \citep{tran2021epoch, blunt2023_overfitting}. Nonetheless, the distinctive presence of transit timing variations in TOI-1224 c provides a promising avenue for future mass constraints. This opportunity is poised to be further explored during the upcoming re-observation of the system in \tess\ Cycle 7.

\begin{longrotatetable}
\begin{deluxetable*}{l  c  c  c  c  c  c  c  c  c  c  c  c c c c c c c c c r}
\tabletypesize{\scriptsize} 
\tablecaption{MELANGE-5 Members \label{tab:toi1224_friends}}
\tablehead{
\colhead{Gaia DR3}&
\colhead{TIC}&
\colhead{$\alpha$}&
\colhead{$\delta$}&
\colhead{RV}&
\colhead{$\sigma_{RV}$}&
\colhead{RV}&
\colhead{EqW Li}&
\colhead{$\sigma_{EqW Li}$}&
\colhead{Li}&
\colhead{$P_{rot}$}&
\colhead{Rotation}\\
\colhead{}&
\colhead{}&
\colhead{J2016.0}&
\colhead{J2016.0}&
\colhead{(km s$^{-1})$}&
\colhead{(km s$^{-1})$}&
\colhead{Source}&
\colhead{m$\AA$}&
\colhead{m$\AA$}&
\colhead{Source}&
\colhead{(days)}&
\colhead{Quality}&
}
\startdata
4620009665047355520 & 299798795 & 37.19798 & -80.8991 & 14.63 & 1.92 & Gaia DR3 & $\cdots$ & $\cdots$ & $\cdots$ & 1.23 & 0.0 \\
5194732707942826496 & 278290227 & 112.69671 & -82.43623 & 32.04 & 0.31 & Gaia DR3 & $\cdots$ & $\cdots$ & $\cdots$ & 6.74 & 1.0 \\
4692269642240299776 & 52242947 & 19.93763 & -68.71285 & $\cdots$ & $\cdots$ & None & $\cdots$ & $\cdots$ & $\cdots$ & 0.44 & 0.0 \\
6357002092906608512 & 287148041 & 338.14076 & -76.93951 & -10.44 & 4.52 & Gaia DR3 & $\cdots$ & $\cdots$ & $\cdots$ & 1.21 & 0.0 \\
5765467482863792768 & 290492468 & 241.10938 & -87.40711 & 5.19 & 3.81 & Gaia DR3 & $\cdots$ & $\cdots$ & $\cdots$ & 1.51 & 3.0 \\
6369133825033348224 & 271577579 & 319.40123 & -74.56481 & 1.64 & 2.94 & Gaia DR3 & $\cdots$ & $\cdots$ & $\cdots$ & 0.71 & 0.0 \\
6387813329294043520 & 402066134 & 352.75559 & -69.08691 & 3.95 & 0.14 & Gaia DR3 & 30.0 & 11.5 & HARPS & $\cdots$ & $\cdots$ \\
6387813363653800448 & 402066136 & 352.76417 & -69.07728 & 3.73 & 0.13 & Gaia DR3 & 98.0 & 14.9 & NRES & 5.22 & 0.0 \\
6361384535800070144 & 1987797195 & 303.43602 & -80.46513 & $\cdots$ & $\cdots$ & None & $\cdots$ & $\cdots$ & $\cdots$ & 3.92 & 0.0 \\
5267405895350690304 & 300015239 & 107.18725 & -70.49845 & $\cdots$ & $\cdots$ & None & $\cdots$ & $\cdots$ & $\cdots$ & 12.77 & 2.0 \\
6369133829329588352 & 271577578 & 319.40049 & -74.56729 & 7.54 & 2.12 & Gaia DR3 & $\cdots$ & $\cdots$ & $\cdots$ & $\cdots$ & $\cdots$ \\
5265069360124273536 & 141756992 & 91.45159 & -73.95555 & $\cdots$ & $\cdots$ & None & $\cdots$ & $\cdots$ & $\cdots$ & 9.57 & 0.0 \\
5771258644968448256 & 418720125 & 210.17054 & -82.22629 & 16.57 & 5.08 & Gaia DR3 & $\cdots$ & $\cdots$ & $\cdots$ & 0.47 & 0.0 \\
4700296214561766144 & 358577029 & 38.53702 & -63.38986 & 29.1 & 2.86 & Gaia DR3 & $\cdots$ & $\cdots$ & $\cdots$ & 3.39 & 0.0 \\
5267405964069463680 & 300015238 & 107.17669 & -70.49661 & 4.97 & 1.17 & Gaia DR3 & $\cdots$ & $\cdots$ & $\cdots$ & $\cdots$ & $\cdots$ \\
5780799416601562496 & 426533052 & 244.59667 & -76.56144 & $\cdots$ & $\cdots$ & None & $\cdots$ & $\cdots$ & $\cdots$ & 0.64 & 0.0 \\
6416270751806848512 & 466680054 & 295.39885 & -72.44675 & 4.6 & 0.14 & Gaia DR3 & 157.0 & 17.85 & NRES & 6.74 & 0.0 \\
4624764296923659648 & 140827419 & 75.25472 & -77.09649 & $\cdots$ & $\cdots$ & None & $\cdots$ & $\cdots$ & $\cdots$ & $\cdots$ & $\cdots$ \\
6420588087293438976 & 409883397 & 293.79044 & -69.97636 & 3.22 & 0.13 & Gaia DR3 & 140.0 & 17.0 & HARPS & 3.92 & $\cdots$ \\
6421013254696711296 & 381982077 & 284.00151 & -70.69675 & 2.72 & 3.18 & Gaia DR3 & $\cdots$ & $\cdots$ & $\cdots$ & 0.26 & 1.0 \\
5196110606467412864 & 323008198 & 125.97079 & -81.97343 & 16.34 & 1.56 & Gaia DR3 & $\cdots$ & $\cdots$ & $\cdots$ & $\cdots$ & 3.0 \\
6396236859672379008 & 327669585 & 330.07931 & -68.00669 & -22.69 & 0.24 & Gaia DR3 & $\cdots$ & $\cdots$ & $\cdots$ & 6.74 & 1.0 \\
6418732764501078784 & 467939527 & 283.81072 & -71.2518 & 40.2 & 0.82 & Gaia DR3 & $\cdots$ & $\cdots$ & $\cdots$ & 7.2 & 1.0 \\
6370853602954758272 & 404348184 & 321.36359 & -73.18337 & 5.52 & 2.91 & Gaia DR3 & $\cdots$ & $\cdots$ & $\cdots$ & 0.67 & 0.0 \\
5482185145257567232 & 150188736 & 93.72865 & -60.65534 & 25.0 & 0.57 & Gaia DR3 & $\cdots$ & $\cdots$ & $\cdots$ & 9.38 & 0.0 \\
5287064510419675776 & 348893965 & 105.90952 & -61.57045 & $\cdots$ & $\cdots$ & None & $\cdots$ & $\cdots$ & $\cdots$ & $\cdots$ & $\cdots$ \\
4757301941103536000 & 309790592 & 80.21079 & -63.41597 & -4.86 & 1.84 & Gaia DR3 & $\cdots$ & $\cdots$ & $\cdots$ & 3.15 & 2.0 \\
4616276930446765824 & 318612083 & 7.57222 & -86.23213 & 14.21 & 0.21 & Gaia DR3 & $\cdots$ & $\cdots$ & $\cdots$ & 6.85 & 0.0 \\
5202659778759682048 & 453442366 & 151.93055 & -77.81114 & 18.94 & 0.35 & Gaia DR3 & $\cdots$ & $\cdots$ & $\cdots$ & 7.48 & 0.0 \\
6356860393345181952 & 317057431 & 336.87475 & -77.7182 & $\cdots$ & $\cdots$ & None & $\cdots$ & $\cdots$ & $\cdots$ & $\cdots$ & $\cdots$ \\
4723828890132788992 & 207234477 & 49.53 & -59.38435 & 3.93 & 1.7 & Gaia DR3 & $\cdots$ & $\cdots$ & $\cdots$ & $\cdots$ & 3.0 \\
6394241555306538240 & 237313246 & 343.79502 & -63.1757 & $\cdots$ & $\cdots$ & None & $\cdots$ & $\cdots$ & $\cdots$ & $\cdots$ & 3.0 \\
6356860393345181824 & 317057430 & 336.88045 & -77.71833 & 4.61 & 0.13 & Gaia DR3 & 55.0 & 12.75 & HARPS & 5.52 & 2.0 \\
5194066717430657792 & 404875401 & 93.78974 & -84.64923 & $\cdots$ & $\cdots$ & None & $\cdots$ & $\cdots$ & $\cdots$ & 0.59 & 0.0 \\
6418406690583059712 & 467934114 & 282.22234 & -72.5607 & 6.4 & 3.43 & Gaia DR3 & $\cdots$ & $\cdots$ & $\cdots$ & 3.67 & 0.0 \\
5270707831848146944 & 453099619 & 116.08608 & -69.5106 & 13.19 & 0.16 & Gaia DR3 & $\cdots$ & $\cdots$ & $\cdots$ & 2.34 & 2.0 \\
6400051752705452544 & 410058419 & 321.05689 & -67.5699 & -7.58 & 4.21 & Gaia DR3 & $\cdots$ & $\cdots$ & $\cdots$ & 0.58 & 0.0 \\
5208911047821367296 & 278518946 & 120.64858 & -79.24193 & 19.71 & 1.53 & Gaia DR3 & $\cdots$ & $\cdots$ & $\cdots$ & 3.12 & 0.0 \\
5223796927729858816 & 303166531 & 137.28797 & -67.99446 & $\cdots$ & $\cdots$ & None & $\cdots$ & $\cdots$ & $\cdots$ & $\cdots$ & $\cdots$ \\
5794951810577913216 & 263792500 & 235.07804 & -71.56834 & 11.79 & 0.15 & Gaia DR3 & -1.0 & 9.95 & HARPS & 20.99 & 2.0 \\
5772759096086560640 & 418807720 & 219.89375 & -80.81085 & 11.75 & 4.94 & Gaia DR3 & $\cdots$ & $\cdots$ & $\cdots$ & 0.66 & 0.0 \\
6487693790194944128 & 234299628 & 358.67472 & -61.587 & 40.4 & 0.31 & Gaia DR3 & $\cdots$ & $\cdots$ & $\cdots$ & 6.24 & 1.0 \\
5800510224109658240 & 260333394 & 225.01435 & -67.55584 & $\cdots$ & $\cdots$ & None & $\cdots$ & $\cdots$ & $\cdots$ & $\cdots$ & $\cdots$ \\
5269380721313848576 & 306931200 & 122.70793 & -71.12608 & 11.74 & 5.48 & Gaia DR3 & $\cdots$ & $\cdots$ & $\cdots$ & 0.64 & 0.0 \\
6401084400282291200 & 372382536 & 317.05673 & -66.38064 & -7.61 & 0.32 & Gaia DR3 & $\cdots$ & $\cdots$ & $\cdots$ & 21.09 & 2.0 \\
5864837701762139776 & 449884234 & 201.30643 & -64.48524 & -4.8 & 0.13 & Gaia DR3 & 0.0 & 10.0 & \citep{Torres2006} & $\cdots$ & 3.0 \\
5864837697361427072 & 449884184 & 201.30292 & -64.47922 & $\cdots$ & $\cdots$ & None & $\cdots$ & $\cdots$ & $\cdots$ & $\cdots$ & $\cdots$ \\
6419363540577655552 & 268750176 & 280.07896 & -71.75748 & $\cdots$ & $\cdots$ & None & $\cdots$ & $\cdots$ & $\cdots$ & $\cdots$ & $\cdots$ \\
4798732668801919360 & 302964354 & 83.34074 & -47.08869 & -14.91 & 0.15 & Gaia DR3 & $\cdots$ & $\cdots$ & $\cdots$ & 9.75 & 0.0 \\
4703216410071320832 & 328011968 & 2.9732 & -69.35424 & $\cdots$ & $\cdots$ & None & $\cdots$ & $\cdots$ & $\cdots$ & $\cdots$ & $\cdots$ \\
5848941306174999424 & 293289676 & 222.03239 & -65.90071 & 42.01 & 0.15 & Gaia DR3 & $\cdots$ & $\cdots$ & $\cdots$ & 6.22 & 1.0 \\
4810476659874420352 & 200363066 & 77.18555 & -45.50202 & 17.54 & 0.28 & Gaia DR3 & $\cdots$ & $\cdots$ & $\cdots$ & 6.97 & 1.0 \\
4836664278068407680 & 101013663 & 55.79118 & -43.53687 & $\cdots$ & $\cdots$ & None & $\cdots$ & $\cdots$ & $\cdots$ & 2.65 & 0.0 \\
6427793736746396672 & 409887354 & 294.23901 & -67.40906 & 8.07 & 0.5 & Gaia DR3 & $\cdots$ & $\cdots$ & $\cdots$ & 12.1 & 2.0 \\
4621748890580775040 & 391949832 & 79.17562 & -80.81982 & 9.14 & 0.9 & Gaia DR3 & $\cdots$ & $\cdots$ & $\cdots$ & 9.6 & 0.0 \\
4814461079558759552 & 200460091 & 71.41374 & -43.89702 & 25.32 & 0.14 & Gaia DR3 & 120.0 & 22.0 & \citep{Torres2006} & 7.99 & 0.0 \\
6638134365589225344 & 456618376 & 282.72856 & -55.75472 & $\cdots$ & $\cdots$ & None & $\cdots$ & $\cdots$ & $\cdots$ & 1.66 & 0.0 \\
6472232212871226240 & 1989423078 & 303.64768 & -54.51718 & $\cdots$ & $\cdots$ & None & $\cdots$ & $\cdots$ & $\cdots$ & 0.57 & 0.0 \\
5779453442570694656 & 407123648 & 234.58773 & -78.05657 & 10.31 & 0.39 & Gaia DR3 & $\cdots$ & $\cdots$ & $\cdots$ & 5.44 & 0.0 \\
5849280505494491264 & 446339621 & 216.94617 & -66.50214 & 7.35 & 2.68 & Gaia DR3 & $\cdots$ & $\cdots$ & $\cdots$ & 0.56 & 0.0 \\
4621599047761938688 & 260995123 & 78.98376 & -81.07174 & $\cdots$ & $\cdots$ & None & -4.0 & 9.8 & HARPS & 9.32 & 2.0 \\
4957985696747200256 & 111267568 & 30.68873 & -40.95115 & 3.23 & 0.13 & Gaia DR3 & 1.0 & 10.05 & HARPS & 20.84 & 2.0 \\
4957985696747200512 & 111267570 & 30.70091 & -40.94715 & 2.57 & 18.86 & Gaia DR3 & $\cdots$ & $\cdots$ & $\cdots$ & $\cdots$ & $\cdots$ \\
4987736729049009024 & 183587880 & 16.89728 & -41.7666 & 15.84 & 0.59 & Gaia DR3 & $\cdots$ & $\cdots$ & $\cdots$ & $\cdots$ & $\cdots$ \\
6428064697644127360 & 325070705 & 293.46493 & -66.8089 & -23.74 & 0.9 & Gaia DR3 & $\cdots$ & $\cdots$ & $\cdots$ & 3.04 & 2.0 \\
5778945223380267648 & 384150620 & 243.6151 & -78.0851 & 9.34 & 0.24 & Gaia DR3 & $\cdots$ & $\cdots$ & $\cdots$ & 6.53 & 0.0 \\
4929504325499059840 & 158565951 & 20.72477 & -49.8718 & -22.66 & 2.94 & Gaia DR3 & $\cdots$ & $\cdots$ & $\cdots$ & 6.76 & 2.0 \\
6475208492064361216 & 100327960 & 307.33502 & -51.49625 & -9.92 & 4.45 & Gaia DR3 & $\cdots$ & $\cdots$ & $\cdots$ & 0.28 & 0.0 \\
5311011770618920832 & 385321596 & 137.79154 & -55.03122 & 26.85 & 0.25 & Gaia DR3 & -1.0 & 9.95 & HARPS & 10.35 & 1.0 \\
4931155036049677312 & 100100909 & 24.36671 & -45.97411 & $\cdots$ & $\cdots$ & None & $\cdots$ & $\cdots$ & $\cdots$ & 0.18 & 1.0 \\
5489939725890739072 & 294154590 & 107.74976 & -56.54963 & 21.5 & 3.9 & \citep{Malo2014a} & $\cdots$ & $\cdots$ & $\cdots$ & 0.19 & 0.0 \\
6473160651658069120 & 201688405 & 300.55649 & -54.43382 & $\cdots$ & $\cdots$ & None & $\cdots$ & $\cdots$ & $\cdots$ & 0.7 & 0.0 \\
5347831082126227712 & 80677028 & 165.98173 & -52.9771 & 21.72 & 0.22 & Gaia DR3 & $\cdots$ & $\cdots$ & $\cdots$ & 10.35 & 1.0 \\
5836174636208941184 & 424969508 & 240.61729 & -56.855 & -10.73 & 0.12 & Gaia DR3 & 76.0 & 13.8 & NRES & 11.04 & 1.0 \\
5489939730188462464 & 766616259 & 107.74932 & -56.54943 & 23.2 & 2.7 & \citep{Malo2014a} & $\cdots$ & $\cdots$ & $\cdots$ & 0.19 & 0.0 \\
6568464361047484288 & 279314834 & 331.68828 & -42.29003 & $\cdots$ & $\cdots$ & None & $\cdots$ & $\cdots$ & $\cdots$ & $\cdots$ & $\cdots$ \\
5895473600324039680 & 412016542 & 214.60387 & -55.47751 & -17.75 & 0.38 & Gaia DR3 & $\cdots$ & $\cdots$ & $\cdots$ & 3.26 & 0.0 \\
4666870820559656320 & 31997119 & 55.57044 & -70.48808 & $\cdots$ & $\cdots$ & None & $\cdots$ & $\cdots$ & $\cdots$ & $\cdots$ & $\cdots$ \\
5924203182025615616 & 213471387 & 257.67238 & -52.11216 & -16.13 & 0.42 & Gaia DR3 & $\cdots$ & $\cdots$ & $\cdots$ & 8.31 & 1.0 \\
5485803711041321216 & 294272204 & 108.51836 & -58.33321 & 22.54 & 5.21 & Gaia DR3 & $\cdots$ & $\cdots$ & $\cdots$ & 0.36 & 0.0 \\
6065158377962553344 & 241655722 & 207.77287 & -53.73532 & $\cdots$ & $\cdots$ & None & -3.0 & 9.85 & HARPS & 0.33 & 0.0 \\
6382707266014366976 & 410150587 & 340.0304 & -71.20608 & 8.12 & 7.64 & Gaia DR3 & $\cdots$ & $\cdots$ & $\cdots$ & 1.52 & 0.0 \\
5843122587538897920 & 329545966 & 192.05181 & -70.64542 & -2.55 & 0.39 & Gaia DR3 & $\cdots$ & $\cdots$ & $\cdots$ & 8.11 & 0.0 \\
5228172606059735936 & 295667714 & 171.55456 & -71.657 & 4.9 & 2.15 & Gaia DR3 & $\cdots$ & $\cdots$ & $\cdots$ & 0.15 & 1.0 \\
5236626338671861760 & 280859868 & 173.26805 & -66.93342 & $\cdots$ & $\cdots$ & None & $\cdots$ & $\cdots$ & $\cdots$ & 0.18 & 0.0 \\
6471472278538182656 & 100662048 & 310.24591 & -52.86164 & 1.09 & 0.41 & Gaia DR3 & $\cdots$ & $\cdots$ & $\cdots$ & 8.66 & 1.0 \\
5808631010846251264 & 1206158616 & 247.96385 & -69.8877 & 6.8 & 0.6 & Gaia DR3 & $\cdots$ & $\cdots$ & $\cdots$ & $\cdots$ & $\cdots$ \\
6516607269516143488 & 121475280 & 343.01097 & -45.92732 & -20.2 & 1.5 & Gaia DR3 & $\cdots$ & $\cdots$ & $\cdots$ & $\cdots$ & 3.0 \\
6445000062927795456 & 411955564 & 290.6827 & -61.73302 & -8.03 & 1.61 & Gaia DR3 & $\cdots$ & $\cdots$ & $\cdots$ & 3.67 & 2.0 \\
6473378904715812608 & 101843377 & 304.09703 & -53.81459 & $\cdots$ & $\cdots$ & None & $\cdots$ & $\cdots$ & $\cdots$ & $\cdots$ & 3.0 \\
4807503271195302272 & 192790473 & 82.11748 & -39.37089 & 27.62 & 0.18 & Gaia DR3 & 90.0 & 14.5 & NRES & 8.51 & 0.0 \\
4807503030677131392 & 192790476 & 82.14335 & -39.37304 & 27.14 & 0.2 & Gaia DR3 & $\cdots$ & $\cdots$ & $\cdots$ & 10.49 & 0.0 \\
5784341905625638400 & 418250020 & 195.98006 & -80.72364 & $\cdots$ & $\cdots$ & None & $\cdots$ & $\cdots$ & $\cdots$ & 2.32 & 0.0 \\
5799237986096386176 & 261517097 & 227.10181 & -69.9266 & 38.45 & 1.52 & Gaia DR3 & $\cdots$ & $\cdots$ & $\cdots$ & 1.61 & 0.0 \\
5212667151341318528 & 177255827 & 101.67769 & -77.00024 & 18.84 & 2.27 & Gaia DR3 & $\cdots$ & $\cdots$ & $\cdots$ & 6.16 & 0.0 \\
5949144267577982720 & 30087015 & 264.89203 & -45.98383 & $\cdots$ & $\cdots$ & None & $\cdots$ & $\cdots$ & $\cdots$ & 1.72 & 2.0 \\
5830329666688142976 & 350798917 & 246.86026 & -61.30937 & -9.22 & 13.99 & Gaia DR3 & $\cdots$ & $\cdots$ & $\cdots$ & $\cdots$ & $\cdots$ \\
5808065209030685568 & 301096165 & 254.69636 & -70.00533 & $\cdots$ & $\cdots$ & None & $\cdots$ & $\cdots$ & $\cdots$ & 0.12 & 2.0 \\
5374565879145559424 & 162434199 & 169.61116 & -47.36567 & $\cdots$ & $\cdots$ & None & $\cdots$ & $\cdots$ & $\cdots$ & $\cdots$ & 3.0 \\
5844099503627340416 & 341055901 & 202.44755 & -69.87024 & 1.72 & 2.13 & Gaia DR3 & $\cdots$ & $\cdots$ & $\cdots$ & 10.02 & 2.0 \\
6637448957528512128 & 1816965664 & 281.35069 & -57.01822 & -9.7 & 4.39 & Gaia DR3 & $\cdots$ & $\cdots$ & $\cdots$ & $\cdots$ & 3.0 \\
5808065209026675584 & 301096163 & 254.68868 & -70.00536 & $\cdots$ & $\cdots$ & None & $\cdots$ & $\cdots$ & $\cdots$ & $\cdots$ & $\cdots$ \\
4951967794731342464 & 91641297 & 40.1676 & -39.8555 & -8.13 & 0.12 & Gaia DR3 & 0.0 & 10.0 & HARPS & $\cdots$ & $\cdots$ \\
6638822694226428672 & 421985503 & 289.09009 & -58.13617 & $\cdots$ & $\cdots$ & None & $\cdots$ & $\cdots$ & $\cdots$ & 0.82 & 2.0 \\
5226155861208509440 & 454295341 & 166.98254 & -74.3516 & $\cdots$ & $\cdots$ & None & $\cdots$ & $\cdots$ & $\cdots$ & $\cdots$ & $\cdots$ \\
6361144915282549504 & 398383318 & 284.15387 & -79.71561 & -6.45 & 0.44 & Gaia DR3 & $\cdots$ & $\cdots$ & $\cdots$ & 14.47 & 0.0 \\
5544880780046222464 & 144355325 & 121.29768 & -35.36725 & 27.98 & 2.24 & Gaia DR3 & $\cdots$ & $\cdots$ & $\cdots$ & 0.59 & 0.0 \\
6098895724014519936 & 129228711 & 220.85415 & -44.4849 & 3.52 & 0.31 & Gaia DR3 & $\cdots$ & $\cdots$ & $\cdots$ & 2.45 & 1.0 \\
4812461514583738752 & 161637624 & 75.96187 & -43.47932 & 25.74 & 0.55 & Gaia DR3 & $\cdots$ & $\cdots$ & $\cdots$ & 0.19 & 0.0 \\
6059555129283005824 & 271508756 & 186.82305 & -57.96319 & 23.82 & 1.44 & Gaia DR3 & $\cdots$ & $\cdots$ & $\cdots$ & 5.22 & 1.0 \\
5307887302169199360 & 441850362 & 145.62191 & -55.57533 & -43.94 & 39.05 & Gaia DR3 & $\cdots$ & $\cdots$ & $\cdots$ & 20.58 & 2.0 \\
5499416283468435712 & 260131643 & 92.61806 & -54.92268 & 73.45 & 0.53 & Gaia DR3 & $\cdots$ & $\cdots$ & $\cdots$ & 2.99 & 2.0 \\
5201153963224913152 & 454291823 & 166.88595 & -77.47413 & $\cdots$ & $\cdots$ & None & $\cdots$ & $\cdots$ & $\cdots$ & 0.68 & 0.0 \\
5212473847749416576 & 177352809 & 104.26012 & -77.16416 & $\cdots$ & $\cdots$ & None & $\cdots$ & $\cdots$ & $\cdots$ & $\cdots$ & $\cdots$ \\
5593952946066050304 & 151006728 & 118.33462 & -33.4419 & 9.19 & 0.14 & Gaia DR3 & $\cdots$ & $\cdots$ & $\cdots$ & 15.4 & 2.0 \\
5440696864990921984 & 105882667 & 157.10331 & -39.98972 & 25.42 & 2.16 & Gaia DR3 & $\cdots$ & $\cdots$ & $\cdots$ & 0.64 & 0.0 \\
5910453793394324096 & 306275682 & 264.70749 & -62.8155 & -0.81 & 12.06 & Gaia DR3 & $\cdots$ & $\cdots$ & $\cdots$ & 0.32 & 1.0 \\
5846665969920173440 & 449041310 & 215.56929 & -69.08548 & -5.3 & 0.17 & Gaia DR3 & $\cdots$ & $\cdots$ & $\cdots$ & 6.53 & 1.0 \\
5575275335880709760 & 393452839 & 94.60879 & -36.9711 & 102.78 & 0.52 & Gaia DR3 & $\cdots$ & $\cdots$ & $\cdots$ & 6.69 & 1.0 \\
4663081800416386816 & 55559834 & 73.18483 & -65.2817 & 68.23 & 4.84 & Gaia DR3 & $\cdots$ & $\cdots$ & $\cdots$ & 1.09 & 0.0 \\
5203361404618057984 & 453220734 & 146.30928 & -77.88724 & $\cdots$ & $\cdots$ & None & $\cdots$ & $\cdots$ & $\cdots$ & $\cdots$ & $\cdots$ \\
5816991972140661760 & 304074982 & 258.74004 & -64.37733 & -20.63 & 5.67 & Gaia DR3 & $\cdots$ & $\cdots$ & $\cdots$ & 0.6 & 0.0 \\
5549662762266400256 & 219164682 & 90.62517 & -51.81968 & 19.3 & 6.65 & Gaia DR3 & $\cdots$ & $\cdots$ & $\cdots$ & 0.57 & 0.0 \\
4812970180447752832 & 200340962 & 76.89603 & -42.0083 & $\cdots$ & $\cdots$ & None & $\cdots$ & $\cdots$ & $\cdots$ & $\cdots$ & $\cdots$ \\
5793819146109422848 & 402682344 & 223.94074 & -73.16681 & 10.5 & 0.13 & Gaia DR3 & 31.0 & 11.55 & NRES & 1.84 & 0.0 \\
4856812863008160256 & 165213021 & 56.50662 & -36.96483 & 19.13 & 0.26 & Gaia DR3 & $\cdots$ & $\cdots$ & $\cdots$ & 6.97 & 0.0 \\
6810820947424835712 & 209375295 & 325.44577 & -27.08304 & -10.85 & 1.69 & Gaia DR3 & $\cdots$ & $\cdots$ & $\cdots$ & $\cdots$ & $\cdots$ \\
5609768282659317376 & 64540162 & 106.92959 & -27.71274 & -35.48 & 6.64 & Gaia DR3 & $\cdots$ & $\cdots$ & $\cdots$ & 0.47 & 1.0 \\
5814435195287770368 & 293075283 & 254.23528 & -68.48423 & $\cdots$ & $\cdots$ & None & $\cdots$ & $\cdots$ & $\cdots$ & 0.44 & 0.0 \\
5880617480245383168 & 416196764 & 223.40451 & -58.28627 & -3.73 & 3.19 & Gaia DR3 & $\cdots$ & $\cdots$ & $\cdots$ & $\cdots$ & 3.0 \\
5837977594661266688 & 360405963 & 188.37251 & -75.38639 & 16.18 & 0.28 & Gaia DR3 & 127.0 & 16.35 & HARPS & 5.44 & 0.0 \\
4655505684288949888 & 30187438 & 73.81168 & -68.64083 & 18.15 & 4.4 & Gaia DR3 & $\cdots$ & $\cdots$ & $\cdots$ & $\cdots$ & 3.0 \\
5348350154694000000 & 91474396 & 170.94489 & -52.96087 & 4.13 & 0.41 & Gaia DR3 & $\cdots$ & $\cdots$ & $\cdots$ & 5.67 & 0.0 \\
2391670474561181440 & 9210746 & 352.55748 & -20.39187 & $\cdots$ & $\cdots$ & None & 0.0 & 10.0 & HARPS & 7.63 & 2.0 \\
5601615988055197952 & 127643940 & 117.24332 & -27.08892 & $\cdots$ & $\cdots$ & None & $\cdots$ & $\cdots$ & $\cdots$ & $\cdots$ & $\cdots$ \\
5795711199462066816 & 263191217 & 234.12269 & -71.56552 & 3.63 & 7.83 & Gaia DR3 & $\cdots$ & $\cdots$ & $\cdots$ & 0.71 & 2.0 \\
5601266725623628416 & 776524223 & 119.80877 & -26.62498 & $\cdots$ & $\cdots$ & None & $\cdots$ & $\cdots$ & $\cdots$ & $\cdots$ & $\cdots$ \\
5974322293543549440 & 200105910 & 261.02843 & -37.46552 & -9.68 & 1.85 & Gaia DR3 & $\cdots$ & $\cdots$ & $\cdots$ & 8.01 & 2.0 \\
6474015178351213824 & 100273626 & 307.12998 & -53.61092 & -4.68 & 1.51 & Gaia DR3 & $\cdots$ & $\cdots$ & $\cdots$ & 9.02 & 0.0 \\
6443749334090478976 & 1988709541 & 300.27412 & -60.37428 & -4.17 & 3.11 & Gaia DR3 & $\cdots$ & $\cdots$ & $\cdots$ & 0.82 & 0.0 \\
6035956895316486656 & 1255391268 & 240.48347 & -33.95344 & -27.31 & 1.19 & Gaia DR3 & $\cdots$ & $\cdots$ & $\cdots$ & 2.96 & 2.0 \\
6035956925371144832 & 1255391255 & 240.4837 & -33.95334 & -27.25 & 0.59 & Gaia DR3 & $\cdots$ & $\cdots$ & $\cdots$ & 2.96 & 2.0 \\
5601950690575573120 & 776770284 & 118.47594 & -26.86299 & $\cdots$ & $\cdots$ & None & $\cdots$ & $\cdots$ & $\cdots$ & $\cdots$ & $\cdots$ \\
5920596066603197056 & 76392634 & 269.69148 & -55.02187 & $\cdots$ & $\cdots$ & None & $\cdots$ & $\cdots$ & $\cdots$ & $\cdots$ & $\cdots$ \\
5782666490422418816 & 357665438 & 188.51063 & -83.53518 & $\cdots$ & $\cdots$ & None & $\cdots$ & $\cdots$ & $\cdots$ & $\cdots$ & $\cdots$ \\
5614014978832904832 & 779943454 & 115.97402 & -25.43424 & $\cdots$ & $\cdots$ & None & $\cdots$ & $\cdots$ & $\cdots$ & $\cdots$ & $\cdots$ \\
4988735051246293504 & 183596242 & 18.36797 & -38.35104 & 14.3 & 0.5 & \citep{Malo2014a} & $\cdots$ & $\cdots$ & $\cdots$ & $\cdots$ & $\cdots$ \\
2395031273585836288 & 434103018 & 353.20741 & -16.84662 & -1.14 & 0.13 & Gaia DR3 & 0.0 & 10.0 & HARPS & 10.09 & 0.0 \\
6635630910753507968 & 119337122 & 276.29233 & -59.01216 & $\cdots$ & $\cdots$ & None & $\cdots$ & $\cdots$ & $\cdots$ & $\cdots$ & $\cdots$ \\
4762827532481570432 & 382066186 & 79.61502 & -58.47309 & 6.98 & 0.57 & Gaia DR3 & $\cdots$ & $\cdots$ & $\cdots$ & 2.38 & 2.0 \\
5811333301184104320 & 293802706 & 263.43478 & -69.31955 & 44.38 & 0.59 & Gaia DR3 & $\cdots$ & $\cdots$ & $\cdots$ & 13.11 & 2.0 \\
2966316109264052096 & 160301040 & 88.62609 & -19.70445 & 42.46 & 0.12 & Gaia DR3 & 8.0 & 10.4 & HARPS & 5.54 & 0.0 \\
5204815332648173440 & 453595277 & 155.58719 & -75.16585 & 19.48 & 0.47 & Gaia DR3 & $\cdots$ & $\cdots$ & $\cdots$ & 7.9 & 0.0 \\
6566592270703002496 & 88273820 & 333.23474 & -46.46388 & 24.1 & 3.85 & Gaia DR3 & $\cdots$ & $\cdots$ & $\cdots$ & $\cdots$ & 3.0 \\
5455707157211784832 & 188043641 & 160.8667 & -29.06451 & 22.57 & 0.13 & Gaia DR3 & 156.0 & 17.8 & HARPS & 6.9 & 0.0 \\
5825336750035914624 & 455606289 & 229.65218 & -64.46337 & 6.75 & 0.23 & Gaia DR3 & $\cdots$ & $\cdots$ & $\cdots$ & 8.91 & 0.0 \\
6788656957673130112 & 289934729 & 318.16972 & -29.37232 & $\cdots$ & $\cdots$ & None & $\cdots$ & $\cdots$ & $\cdots$ & $\cdots$ & $\cdots$ \\
5217796343023874432 & 452475180 & 139.65738 & -73.8369 & $\cdots$ & $\cdots$ & None & $\cdots$ & $\cdots$ & $\cdots$ & 0.67 & 0.0 \\
5494935911023714944 & 260267208 & 93.91221 & -57.70134 & 27.35 & 0.23 & Gaia DR3 & 140.0 & 17.0 & HARPS & 3.68 & 0.0 \\
\enddata
\end{deluxetable*}
\end{longrotatetable}

\begin{deluxetable*} {lccccccccr} 
\tabletypesize{\footnotesize} 
\tablecaption{Young ($<$ 1 Gyr) Multi-planetary Systems \label{tab:young_multis}}
\tablecolumns{9}
\tablewidth{0pt}
 \tablehead{
\colhead{Star Name} & 
\colhead{Cluster/} & 
\colhead{Age} &
\colhead{R$_{*}$} & 
\colhead{T$_{eff}$} &
\colhead{Planet Name} &
\colhead{R$_{p}$}  & 
\colhead{T$_{eq}$} & 
\colhead{Period} & 
\colhead{Reference} \\
\colhead{} & 
\colhead{Association} & 
\colhead{(Myr)} & 
\colhead{(R$_{\odot}$)} & 
\colhead{(K)} & 
\colhead{} & 
\colhead{(R$_{\oplus}$)} & 
\colhead{(K)} & 
\colhead{(days)} & 
\colhead{}}
\startdata
\multirow{2}{*}{K2-264} & \multirow{2}{*}{Praesepe} &  \multirow{2}{*}{700} & \multirow{2}{*}{0.47} & \multirow{2}{*}{3660} & b & 2.23 & 496 & 5.84 & \multirow{2}{*}{\citet{Livingston2019MNRAS.484....8L}} \\ 
 &  &  &  &  & c & 2.67 & 331 & 19.66  \\
\hline
\multirow{3}{*}{K2-136} & \multirow{3}{*}{Hyades} &  \multirow{3}{*}{700} & \multirow{3}{*}{0.68} & \multirow{3}{*}{4500} & b & 1.01 & 610 & 7.98 & \multirow{3}{*}{\citet{Mayo2023AJ....165..235M}}  \\ 
&  &  &  &  & c & 3.00 & 470 & 17.31 \\
&  &  &  &  & d & 1.57 & 420 & 25.58 \\
\hline
\multirow{3}{*}{HD 63433} & \multirow{3}{*}{Ursa Major} &  \multirow{3}{*}{414} & \multirow{3}{*}{0.91} & \multirow{3}{*}{5640} & b & 2.11 & 968 & 7.11 & \multirow{3}{*}{\citet{Capistrant2024AJ....167...54C}} \\
&  &  &  & & c & 2.52 & 679 & 20.54  \\
&  &  &  & & d & 1.07 & 1040 & 4.209 \\
\hline
\multirow{3}{*}{TOI-2076} & \multirow{3}{*}{Crius 224} &  \multirow{3}{*}{340} & \multirow{3}{*}{0.77} & \multirow{3}{*}{5200} & b & 2.52 & 797 & 10.36 & \multirow{3}{*} {\citet{osborn2022uncovering}} \\ 
&  &  &  & & c & 3.50 & 623 & 21.02 \\
&  &  &  & & d & 3.23 & 530 & 35.13 \\
\hline
\multirow{2}{*}{TOI-1224} & \multirow{2}{*}{MELANGE-5} &  \multirow{2}{*}{\age\,} & \multirow{2}{*}{0.44} & \multirow{2}{*}{3326} & b & 2.10 & 540 & 4.18 & \multirow{2}{*}{This Work}  \\
&  &  &  & & c & 2.88 & 332 & 17.95  \\ 
\hline
\multirow{3}{*}{TOI-451} & \multirow{3}{*}{Pisces–Eridanus} &  \multirow{3}{*}{120} & \multirow{3}{*}{0.88} & \multirow{3}{*}{5550} & b & 1.91 & 1491 & 1.86 & \multirow{3}{*}{\citet{Newton2021AJ....161...65N}} \\ 
&  &  &  &  & c & 3.10 & 875 & 9.19 \\ 
&  &  &  &  & d & 4.07 & 722 & 16.36 \\ 
\hline
\multirow{2}{*}{HD 109833} & \multirow{2}{*}{MELANGE-4} &  \multirow{2}{*}{27} & \multirow{2}{*}{1.00} & \multirow{2}{*}{5881}  & b & 2.89 & 811 & 9.19 & \multirow{2}{*}{ \citet{Wood2023}} \\ 
&  &  &  &  & c & 2.59 & 757 & 13.90 \\ 
\hline
\multirow{4}{*}{V1298 Tau} & \multirow{4}{*}{Taurus} &  \multirow{4}{*}{20} & \multirow{4}{*}{1.33} & \multirow{4}{*}{5050} & b & 9.53 & 677 & 24.14 & \multirow{4}{*}{\citet{Feinstein2022ApJ...925L...2F}} \\ 
&  &  &  &  &  c & 5.05 & 968 & 8.24  \\ 
&  &  &  &  &  d & 6.13 & 845 & 12.40 \\ 
&  &  &  &  &   e & 9.94 & 492 & 44.17  \\
\hline
\multirow{2}{*}{AU Mic} & \multirow{2}{*}{Beta Pictoris} &  \multirow{2}{*}{18} & \multirow{2}{*}{0.82} & \multirow{2}{*}{3665} & b & 4.19 & 593 & 8.46 & \multirow{2}{*}{\citet{Donati2023MNRAS.525..455D}} \\ 
&  &  &  &  &  c & 2.79 & 454 & 18.86 \\ 
\enddata
\end{deluxetable*}

\acknowledgements

We thank the anonymous referee for their careful reading and thoughtful comments on the manuscript.

The authors wish to acknowledge Wally, Penny, Bandit, and Halee for their tireless endeavors towards research and exploration.

PCT was supported by NSF Graduate Research Fellowship (DGE-1650116), the NC Space Grant Graduate Research Fellowship, the Zonta International Amelia Earhart Fellowship, and the Jack Kent Cooke Foundation Graduate Scholarship. AWM was supported by grants from the NSF CAREER program (AST-2143763) and NASA's exoplanet research program (XRP 80NSSC21K0393). MGB was supported by NSF Graduate Research Fellowship (DGE-2040435) and the NC Space Grant Graduate Research Fellowship.  F.J.P acknowledges financial support from the grant CEX2021-001131-S funded by MCIN/AEI/ 10.13039/501100011033 and through projects PID2019-109522GB-C52 and PID2022-137241NB-C43. 

This paper made use of data collected by the TESS mission and are publicly available from the Mikulski Archive for Space Telescopes (MAST) operated by the Space Telescope Science Institute (STScI). 
 
We acknowledge the use of public TESS data from pipelines at the TESS Science Office and at the TESS Science Processing Operations Center. 
 
Resources supporting this work were provided by the NASA High-End Computing (HEC) Program through the NASA Advanced Supercomputing (NAS) Division at Ames Research Center for the production of the SPOC data products.

This work has made use of data from the European Space Agency (ESA) mission {\it Gaia} (\url{https://www.cosmos.esa.int/gaia}), processed by the {\it Gaia} Data Processing and Analysis Consortium (DPAC,
\url{https://www.cosmos.esa.int/web/gaia/dpac/consortium}). Funding for the DPAC has been provided by national institutions, in particular the institutions
participating in the {\it Gaia} Multilateral Agreement.

This work makes use of observations from the ASTEP
telescope. ASTEP benefited from the support of the
French and Italian polar agencies IPEV and PNRA in
the framework of the Concordia station program, from
OCA, INSU, Idex UCAJEDI (ANR- 15-IDEX-01) and
ESA through the Science Faculty of the European Space Research and Technology Centre (ESTEC).
This research also received funding from the European Research Council (ERC) under the European Union's Horizon 2020 research and innovation programme (grant agreement n$^\circ$ 803193/BEBOP), and from the Science and Technology Facilities Council (STFC; grant n$^\circ$ ST/S00193X/1).

This work makes use of observations from the LCOGT network. Part of the LCOGT telescope time was granted by NOIRLab through the Mid-Scale Innovations Program (MSIP). MSIP is funded by NSF.

The research leading to these results has received funding from the ARC grant for Concerted Research Actions, financed by the Wallonia-Brussels Federation. TRAPPIST is funded by the Belgian Fund for Scientific Research (Fond National de la Recherche Scientifique, FNRS) under the grant PDR T.0120.21. MG is F.R.S.-FNRS Research Director and EJ is F.R.S.-FNRS Senior Research Associate. Observations were carried out from ESO La Silla Observatory.

The postdoctoral fellowship of KB is funded by F.R.S.-FNRS grant T.0109.20 and by the Francqui Foundation. KAC and acknowledges support from the TESS mission via subaward s3449 from MIT. MK acknowledges support from the MIT Kavli Institute as a Juan Carlos Torres Fellow.

 This publication benefits from the support of the French Community of Belgium in the context of the FRIA Doctoral Grant awarded to MT.

This research has made use of the NASA Exoplanet Archive, which is operated by the California Institute of Technology, under contract with the National Aeronautics and Space Administration under the Exoplanet Exploration Program.

\software{\texttt{LDTK} \citep{2015MNRAS.453.3821P}, \texttt{MISTTBORN}, \texttt{emcee} \citep{Foreman-Mackey2013} , \texttt{corner.py} \citep{foreman2016corner}, \texttt{celerite} \citep{celerite}, \texttt{matplotlib} \citep{hunter2007matplotlib}, \texttt{batman} \citep{Kreidberg2015}, \texttt{Astropy} \citep{Astropy2013, AstropyCollaboration2018, Astropy2022}, \texttt{numpy} \citep{Harris2020}, AstroImageJ \citep{Collins17}, TAPIR \citep{Jensen:2013}, \texttt{juliet} \citep{espinoza2019juliet}, \texttt{dynesty} \citep{sergey_koposov_2023_7600689, speagle2020dynesty}}

\facilities{\textit{TESS}, \textit{Gaia}, ASTEP, LCOGT, TRAPPIST-South, Exoplanet Archive}
\clearpage

\bibliography{bib.bib,pg.bib}

\end{document}